\documentclass[sigconf]{acmart}

\AtBeginDocument{%
  \providecommand\BibTeX{{%
    \normalfont B\kern-0.5em{\scshape i\kern-0.25em b}\kern-0.8em\TeX}}}

\setcopyright{acmlicensed}
\copyrightyear{2026}
\acmYear{2026}
\acmDOI{XXXXXXX.XXXXXXX}

\acmConference[SIGMOD '26]{the 2026 International Conference
on Management of Data}{May 31--June 05,
  2026}{Bengaluru, India}
\acmISBN{978-1-4503-XXXX-X/18/06}

\usepackage{graphicx}
\usepackage{subfigure}
\usepackage{multirow}
\usepackage{amsthm}
\usepackage[explicit]{titlesec}
\usepackage[noend, linesnumbered, ruled, lined, nofillcomment]{algorithm2e}
\usepackage{hyperref}
\usepackage{balance}
\usepackage{color}
\usepackage{pifont}
\usepackage{amsthm} 
\usepackage{amsmath}

\usepackage{flushend}
\usepackage{bm}
\usepackage{array}

\begin{document}

\title[Scalable Graph Indexing using GPUs for Approximate Nearest Neighbor Search]
{\texorpdfstring{Scalable Graph Indexing using GPUs \\for Approximate Nearest Neighbor Search}{Scalable Graph Indexing using GPUs for Approximate Nearest Neighbor Search}}


\author{Zhonggen Li}
\affiliation{%
  \institution{Zhejiang University}
}
\email{zgli@zju.edu.cn}

\author{Xiangyu Ke}
\affiliation{%
  \institution{Zhejiang University}
}
\email{xiangyu.ke@zju.edu.cn}

\author{Yifan Zhu}
\affiliation{%
  \institution{Zhejiang University}
}
\email{xtf_z@zju.edu.cn}

\author{Bocheng Yu}
\affiliation{%
  \institution{Zhejiang University}
}
\email{yubc@zju.edu.cn}

\author{Baihua Zheng}
\affiliation{%
  \institution{Singapore Management University}
}
\email{bhzheng@smu.edu.sg}

\author{Yunjun Gao}
\affiliation{%
  \institution{Zhejiang University}
}
\email{gaoyj@zju.edu.cn}

\renewcommand{\shortauthors}{Zhonggen Li, Xiangyu Ke, Yifan Zhu, Bocheng Yu, Baihua Zheng, and Yunjun Gao}

\begin{abstract}
Approximate nearest neighbor search (ANNS) in high-dimensional vector spaces has a wide range of real-world applications. Numerous methods have been proposed to handle ANNS efficiently, while graph-based indexes have gained prominence due to their high accuracy and efficiency. However, the indexing overhead of graph-based indexes remains substantial. With exponential growth in data volume and increasing demands for dynamic index adjustments, this overhead continues to escalate, posing a critical challenge. 

In this paper, we introduce {\sf Tagore}, a fas\textbf{\underline{T}} library \textbf{\underline{a}}ccelerated by \textbf{\underline{G}}PUs f\textbf{\underline{or}} graph ind\textbf{\underline{e}}xing, which has powerful capabilities of constructing refinement-based graph indexes such as NSG and Vamana. We first introduce GNN-Descent, a GPU-specific algorithm for efficient \textit{k}-Nearest Neighbor (\textit{k}-NN) graph initialization. GNN-Descent speeds up the similarity comparison by a two-phase descent procedure and enables highly parallelized neighbor updates. Next, aiming to support various \textit{k}-NN graph pruning strategies, we formulate a universal pruning procedure termed CFS and devise two generalized GPU kernels for parallel processing complex dependencies in neighbor relationships. For large-scale datasets exceeding GPU memory capacity, we propose an asynchronous GPU-CPU-disk indexing framework with a cluster-aware caching mechanism to minimize the I/O pressure on the disk. Extensive experiments on 7 real-world datasets exhibit that {\sf Tagore} achieves 1.32$\times$ to 112.79$\times$ speedup while maintaining the index quality. 
\end{abstract}

\maketitle

\vspace{-0.8mm}
\section{Introduction}
\label{sec:intro}

\begin{figure}[tp]
    \centering
    \includegraphics[width=1\linewidth]{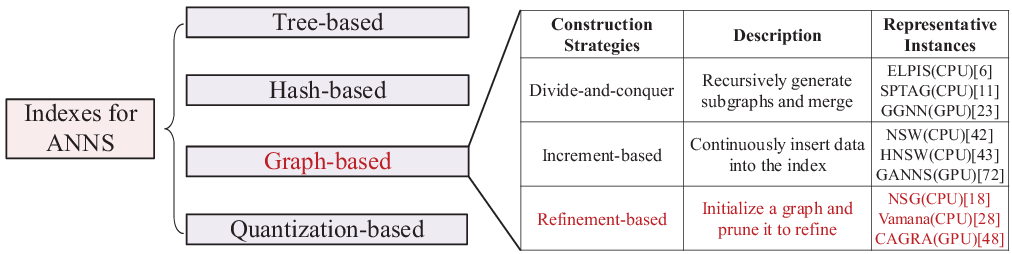}
    \caption{{Categories of Index construction methods for ANNS. {\sf Tagore} focuses on the refinement-based methods.}}
    \vspace{-6mm}
    \label{fig:targeted_problem}
\end{figure}

Approximate nearest neighbor search (ANNS) is a fundamental component in database systems~\cite{pan2024survey,gao2024rabitq,su2024vexless,gao2023high}, with numerous real-world applications, including information retrieval~\cite{gou2024semantic,lampropoulos2023adaptive,li2018design}, retrieval-augmented generation~\cite{yu2025aquapipe,ang2024tsgassist,jiang2023chameleon}, and recommendation~\cite{miao2022het,parchas2020fast}. 
To efficiently respond to user queries, various ANNS methods have been explored, including tree-based methods~\cite{zeng2023litehst,echihabi2022hercules,malkov2018efficient}, hash-based methods~\cite{zhang2020continuously,lu2020vhp}, quantization-based methods~\cite{gao2024rabitq,andre2016cache}, and graph-based methods~\cite{peng2023efficient,fu2017nsg,malkov2018efficient}. 
Among these, graph-based methods have gained significant popularity due to remarkable performance in both efficiency and accuracy~\cite{wang2021comprehensive,azizi2023elpis,pan2024survey}. 
{As shown in Figure \ref{fig:targeted_problem}, \textit{divide-and-conquer}, \textit{increment-based}, and \textit{refinement-based} methods are the three primary index construction methods for graph-based index.} 
Recent studies highlight that \textit{refinement-based} indexes (e.g., NSG~\cite{fu2017nsg}, Vamana~\cite{jayaram2019diskann}, NSSG~\cite{fu2021high}) demonstrate superior efficiency in both construction and query ~\cite{wang2021comprehensive,annbenchmark}. 
These methods are widely adopted in commercial vector databases like Milvus~\cite{wang2021milvus}, as well as by companies such as Microsoft and Alibaba~\cite{jayaram2019diskann,singh2021freshdiskann,fu2017nsg}. 

Numerous research initiatives focus on enhancing query efficiency of ANNS, treating index construction as an offline procedure with limited optimization~\cite{gao2023high,li2020improving,lu2021hvs}.
However, efficient index construction has become increasingly crucial for real-world applications~\cite{yang2024revisiting,wang2025accelerating}. 
With the exponential growth of data, the volume of vector data required for retrieval in enterprise-scale databases has surpassed the billion-scale threshold, as seen in Meta's image databases~\cite{johnson2019billion} and Taobao's product vectors~\cite{fu2017nsg}. 
As a result, constructing a graph-based index has become increasingly time-consuming, requiring dozens of hours to build an index on billion-scale datasets~\cite{jayaram2019diskann}. The overhead is insufficient for industrial needs, where nightly index reconstruction is necessary to maintain both accuracy and efficiency.
Furthermore, 
embedding models are frequently updated to align with evolving requirements, and vectors are dynamically inserted or deleted~\cite{singh2021freshdiskann,xu2023spfresh}. 
Although recent studies have investigated in-place index updates, these methods typically suffer a 5\%-10\% accuracy degradation after several update cycles~\cite{wang2025accelerating}. 
Consequently, periodic index rebuilding remains a common strategy to balance efficiency and query accuracy
~\cite{wei2020analyticdb}.

Recent research has proposed various optimized solutions to address the inefficiency of graph index construction on modern CPUs~\cite{wang2025accelerating,gou2025symphonyqg,zhao2023towards,ono2023relative,yang2024revisiting}. 
{However, the construction overhead remains significant on CPUs.}
For instance, constructing \textit{in-memory} indexes for datasets of 1B vectors still requires more than 10 hours, which remains unsatisfactory in terms of efficiency~\cite{wang2025accelerating,gou2025symphonyqg}. 

To further enhance efficiency, some studies have explored the acceleration of graph-based index construction using GPUs~\cite{wang2021fast,yu2022gpu,ootomo2024cagra,groh2022ggnn}. 
While GPU-based methods 
outperform CPU-based methods, the full potential of GPUs remains largely untapped, restricting their widespread applicability. 
For example, GANNS~\cite{yu2022gpu} incurs additional merging overhead when parallelizing the index construction of HNSW on GPUs. \textit{Increment-based} methods (e.g., HNSW and NSW), which require sequentially inserting nodes to ensure index quality, are not inherently well-suited for parallelizing on GPUs. 
In contrast, 
\textit{refinement-based} methods, particularly the core components-initialization (typically via NN-Descent~\cite{dong2011efficient}) and pruning (often using the Relative Neighborhood Graph strategy~\cite{malkov2018efficient})-are naturally parallelizable due to their data decoupling, offering significant potential for acceleration. 
However, existing studies of accelerating \textit{refinement-based} methods using GPUs have failed to achieve satisfactory performance. 
For example, CAGRA~\cite{ootomo2024cagra}  optimizes the pruning process but leaves graph initialization as a bottleneck. 
Additionally, most existing GPU-based methods focus on accelerating the construction of specific indexes or designing customized indexes tailored for GPUs, often lacking support for a broader range of mainstream graph indexes~\cite{ootomo2024cagra,groh2022ggnn,yu2022gpu}. Scalability with large datasets also remains a challenge~\cite{zhang2024fast,khan2024bang,DBLP:conf/fast/TianLTXDL0ZZ025}. 

\begin{figure}
    \centering
    \includegraphics[width=0.43\textwidth]{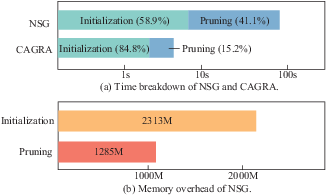}
    \caption{Time breakdown (a) and memory usage (b) of representative refinement-based methods, NSG (CPU) and CAGRA (GPU), on dataset SIFT\protect\footnotemark[1]. }
    
    \label{fig:time_component}
\end{figure}
\footnotetext[1]{{For Figure \ref{fig:time_component}(b), CAGRA shows a similar trend of 2434M memory usage on the GPU.}}

To better understand the inefficiencies in current \textit{refinement-based} methods, we profile the time breakdown and memory consumption of NSG and CAGRA. As shown in Figure~\ref{fig:time_component}(a), NSG exhibits substantial overhead in both the initialization and pruning phases. CAGRA accelerates indexing on the GPU and introduces a GPU-optimized pruning strategy that improves pruning efficiency. However, its initialization phase remains a bottleneck, even when leveraging the state-of-the-art GPU-accelerated NN-Descent algorithm~\cite{wang2021fast}. 
Although GPUs excel at parallel distance computation during NN-Descent, their computational advantages diminish as iteration convergence slows, coupled with persistent overhead from frequent neighbor updates. These updates involve extensive memory access and synchronization locks, reducing GPU parallelism. 
Thus, \textit{slow convergence} and \textit{frequent neighbor updates} are critical factors hindering the efficiency of the initialization phase on GPUs. 

Following initialization, the diversity of pruning strategies complicates the design of \textit{a unified acceleration framework}. 
Moreover, pruning strategies such as NSG~\cite{fu2017nsg}, NSSG~\cite{fu2021high} and Vamana~\cite{jayaram2019diskann} 
require \textit{the serial computation of complex neighbor dependencies}, where neighbors need to be gradually inserted into the neighbor list.
This serial computation proves challenging to parallelize effectively on GPUs. 
Furthermore, as illustrated in Figure~\ref{fig:time_component}(b), the high memory footprint of refinement-based methods makes it difficult for GPUs to handle large-scale datasets due to their limited memory capacity. This often necessitates GPU-CPU-disk coordination, leading to excessive disk I/O, which further degrades performance. Consequently, \textit{efficiently indexing billion-scale datasets with limited GPU memory} and \textit{orchestrating the cooperation between CPU, GPU, and disk} presents another significant challenge. 

To address these challenges, we present {\sf Tagore}, an efficient and scalable GPU-accelerated library that incorporates novel optimizations for initialization, pruning, and large-scale indexing. 
{\sf Tagore} offers user-friendly Python APIs, enabling seamless integration of GPU acceleration into index construction workflows. 
In {\sf Tagore}, we introduce GNN-Descent, a two-phase algorithm designed to accelerate convergence and support lock-free neighbor updates~(\S~\ref{sec:gnn_descent}). In the first phase, adjacent nodes share a batch of samples to exploit GPU parallelism, facilitating efficient neighbor list initialization. The second phase employs node-specific sampling for fine-grained neighbor refinement, reducing computational redundancy. Both phases utilize a lock-free merging strategy to update neighbors, eliminating synchronization overhead. 
To unify the acceleration of various pruning strategies, we propose the CFS (Collect-Filter-Store) framework, which decomposes pruning into three modular stages: 1) \textit{collecting} candidate neighbors, 2) \textit{filtering} based on pruning strategies, and 3) \textit{storing} optimized neighbors~(\S~\ref{subsec:unified_api}). 
Additionally, we develop two specialized GPU kernels for the CFS framework: 1) a \textit{parallel increment kernel} for computing complex neighbor dependencies, and 2) a \textit{parallel balance kernel} for distributing uneven computation tasks across GPU threads~(\S~\ref{subsec:opt_GPU}).
For large-scale datasets, we design an asynchronous GPU-CPU-disk framework that decouples computation, merging, and persistence tasks~(\S~\ref{subsec:asyn_framework}). This framework enables continuous GPU processing of graph construction subtasks, while the CPU asynchronously merges partial indexes and the disk handles large-scale data persistence. To mitigate I/O times, we introduce a clustering-aware caching mechanism that prioritizes merging operations for overlapping node clusters, significantly reducing redundant disk I/O~(\S~\ref{subsec:caching}). 

In summary, this paper makes the following contributions. 
\begin{itemize}
    \item \textit{Two-phase parallelized graph construction.} We introduce a two-phase construction algorithm to accelerate convergence and enable lock-free neighbor updates with high parallelism. 
    \item \textit{Refinement-based pruning acceleration.} Supported by two specialized GPU kernels for efficient serial and unbalanced computation, we propose a unified pipeline to accelerate refinement-based pruning. 
    \item \textit{Asynchronous GPU-CPU-disk framework.} We design an asynchronous GPU-CPU-disk framework and a caching mechanism to index billion-scale datasets within GPU and CPU memory constraints while alleviating disk I/O pressure. 
    \item \textit{Extensive experiments.} We conduct comprehensive experiments demonstrating that {\sf Tagore} outperforms existing methods on 7 real-world datasets, achieving 1.32$\times$-112.79$\times$ speedup. 
\end{itemize}

The rest of this paper is organized as follows. 
Section \ref{sec:related_work} reviews the related studies. 
Section \ref{sec:preliminaries} introduces graph-based index construction and GPU architectures. 
Section \ref{sec:gnn_descent} describes the GNN-Descent algorithm. 
Section \ref{sec:pruning} exhibits the pruning acceleration. 
Section \ref{sec:large_graph} presents the GPU-CPU-disk asynchronous framework. 
Section \ref{sec:experiments} reports the experimental results. 
Section \ref{sec:conclusion} concludes the paper. 
\section{Related Work}
\label{sec:related_work}
In this section, we review the related work about graph-based indexes and indexing acceleration. 
\subsection{Graph-based Indexes}
Graph-based indexes are prominent among various types of ANNS methods due to their superior accuracy and efficiency~\cite{ren2020hm,ngt2018,fu2016efanna,malkov2018efficient}. They model a proximity graph where vertices represent vectors in databases and edges encode the nearest-neighbor relationships. Existing graph-based methods are categorized into three classes based on construction strategies~\cite{wang2021comprehensive}. 

\noindent\textbf{Refinement-based methods} initialize neighbors for each node using randomization~\cite{jayaram2019diskann,kgraph2011}, multiple KD-trees~\cite{fu2016efanna}, or the NN-Descent algorithm~\cite{fu2017nsg,ootomo2024cagra}. While random initialization yields low-quality graphs and KD-trees incur high computational costs~\cite{wang2021comprehensive}, NN-Descent has become a popular initialization method for its balance of efficiency and quality~\cite{fu2017nsg,li2019approximate,fu2021high,ootomo2024cagra}. Subsequent pruning phases refine neighbor distributions to optimize search performance, with strategies like Relative Neighborhood Graph (RNG) in NSG~\cite{fu2017nsg} and rank-based pruning in CAGRA~\cite{ootomo2024cagra} and NGT~\cite{ngt2018}. 

\noindent\textbf{Increment-based methods} construct the index incrementally by inserting vectors into the graph~\cite{malkov2014approximate,malkov2018efficient,ren2020hm}. For each vector, the nearest neighbors are retrieved from the current graph, and edges are attached between new vectors and their nearest neighbors. Moreover, some methods incorporate pruning to determine whether an edge can be established between a vector and its nearest neighbors during construction~\cite{zhao2023towards}. 

\noindent\textbf{Divide-and-conquer} methods partition datasets into clusters and construct local subgraphs for each partition~\cite{sptag18,shimomura2019hgraph,groh2022ggnn,azizi2023elpis,chen2009fast}. They merge the subgraphs to establish a global proximity nearest neighbor graph to ensure high search performance. 

Considering the superior query performance and parallelization-friendly workflow of refinement-based methods, we focus on accelerating the construction of this type of index. 

\subsection{Indexing Acceleration}
\noindent \textbf{Indexing acceleration on CPUs.} Computational inefficiencies in graph indexing have driven research toward reducing computational costs. Existing CPU-based acceleration approaches fall into three categories. The first type aims to redesign the index structure to optimize the time complexity of search~\cite{peng2023efficient, zhao2023towards, ono2023relative,yang2024revisiting}, while the second type attempts to reduce the distance computation overhead via quantization~\cite{gou2025symphonyqg,ngt2018,gao2024rabitq} or statistical estimation~\cite{chen2023finger,yang2024effective}. The third type leverages SIMD instructions of modern CPUs and adjusts the data layout to minimize random memory access~\cite{wang2025accelerating,gou2025symphonyqg}. These methods achieve significant improvement. However, they require substantial resources for large-scale datasets and remain impractical for nightly index reconstruction in production environments. 

\noindent\textbf{Indexing acceleration on GPUs.} 
GPUs’ parallel computing capabilities have been widely exploited for accelerating ANNS queries using graph-based indexes~\cite{huang2024neos,khan2024bang,zhao2020song,ootomo2024cagra,groh2022ggnn,yu2022gpu,DBLP:conf/fast/TianLTXDL0ZZ025,zhu2024gts,liu2024juno}. 
Moreover, some research explores ANNS on other emerging hardware besides GPUs, such as FPGAs~\cite{jiang2023co,jiang2023chameleon,zeng2023df}, SmartSSDs~\cite{tian2024scalable}, and CXL~\cite{jang2023cxl,jang2024bridging}.
Recent work has noticed the inefficiency of indexing and begun leveraging GPUs to optimize index construction~\cite{yu2022gpu,wang2021fast,ootomo2024cagra}. For example, GANNS~\cite{yu2022gpu} and GGNN~\cite{groh2022ggnn} employ a divide-and-conquer strategy, partitioning datasets into subgraphs built in parallel on GPU thread blocks. The subgraphs are finally merged to construct a global graph index. This strategy introduces additional merging overhead, even though on small datasets, resulting in suboptimal performance~\cite{ootomo2024cagra}. 
GNND~\cite{wang2021fast} proposes a GPU-accelerated NN-Descent algorithm, providing a foundation for the acceleration of refinement-based methods. Based on GNND implemented in cuVS~\cite{cuvs2024}, CAGRA~\cite{ootomo2024cagra} introduces a pruning strategy especially for GPUs, achieving state-of-the-art performance in both indexing and search. However, existing GPU methods lack a unified library for accelerating mainstream refinement-based indexes, and the potential of GPUs has not yet been fully realized. 

\noindent\textbf{Out-of-core indexing acceleration.} 
{DiskANN~\cite{jayaram2019diskann} employs a divide-and-conquer strategy to construct the out-of-core index, partitioning the dataset via $k$-means clustering and building local indexes that are later merged into a global index. SPANN~\cite{chen2021spann} adopts an inverted index approach, storing only posting list centroids in memory and utilizing hierarchical balanced clustering for efficient posting lists generation. Starling~\cite{wang2024starling} enhances data locality through block shuffling during index construction. However, these methods rely on the CPU to construct the index, leading to inefficiency and substantial I/O overhead. 
Similar to our proposed method, some out-of-core GNN training frameworks leverage caching techniques to mitigate I/O costs. These methods cache frequently accessed (``hot'') features on the GPU and minimize inter-partition cut edges based on graph connectivity to reduce data transfer overhead, employing strategies such as static caching~\cite{sun2025hyperion}, dynamic caching~\cite{liu2023bgl}, and advanced partitioning~\cite{zheng2022distributed,lin2020pagraph,sun2023legion}. However, their primary challenge lies in identifying hot features and optimizing partitioning, whereas out-of-core indexing primarily requires efficient cluster scheduling for subgraph construction. This distinction renders our problem fundamentally different, with a unique optimization objective. }

\section{Preliminaries}
\label{sec:preliminaries}
In this section, we introduce the background of the refinement-based index construction and GPU architectures. 

\subsection{Refinement-based Indexing}
The construction of refinement-based indexes comprises two phases: \textit{k}-NN graph initialization and graph pruning. Initialization typically employs the NN-Descent algorithm~\cite{dong2011efficient}, which has become the most popular method in KGraph~\cite{kgraph2011}, DPG~\cite{li2019approximate}, NSG~\cite{fu2017nsg}, and CAGRA~\cite{ootomo2024cagra}. Pruning leverages a variety of strategies, including the RNG-based~\cite{fu2017nsg} and rank-based~\cite{ootomo2024cagra} methods. In this subsection, we first outline the NN-Descent algorithm and then summarize key pruning strategies used in refinement-based methods. 

\vspace{-1mm}
\subsubsection{\textbf{NN-Descent.}} 
\label{subsubsection:nn_descent}

The NN-Descent algorithm~\cite{dong2011efficient} is a widely adopted method for constructing approximate \textit{k}-NN graphs. It leverages the principle of ``neighbors are more likely to be neighbors of each other'', and iteratively refines an initial graph where each node is randomly connected to $k$ neighbors. In each iteration, each node introduces its neighbors to each other, allowing them to discover closer neighbors. Specifically, during each iteration, the algorithm (1) samples \textit{new} (previously unsampled) and \textit{old} (already sampled) nodes from the neighbor list of each node, (2) performs a \textit{local join} operation to calculate pairwise distances between \textit{new-new} and \textit{new-old} node pairs, and (3) updates the closer neighbors into the neighbor lists of the corresponding nodes. The new nodes inserted in the current iteration are labeled as \textit{new}, while the previously sampled nodes are labeled as \textit{old}. This process repeats until the approximate \textit{k}-nearest neighbors converge to the desired quality. 

GPU-accelerated implementations optimize NN-Descent by reformulating the \textit{local join} operation as batched matrix multiplication, which GPUs efficiently perform~\cite{wang2021fast,cuvs2024}. However, frequent neighbor list updates lead to significant memory access and synchronization overhead~\cite{wang2021fast}. To mitigate this, existing methods prioritize inserting only the closest neighbor per iteration rather than all closer neighbors identified during the iteration. 

\begin{table}[tbp]
    \renewcommand{\arraystretch}{0.95}
    \centering
    \small
    \caption{Pruning strategies of refinement-based indexes, where $R$ refers to the neighbor list for point $p$. }
    \vspace{-3mm}
    \begin{tabular}{m{1cm}<{\centering}m{3.8cm}<{\centering}m{2.6cm}<{\centering} }
        \toprule
         Index & Pruning strategies & Computing paradigm  \\
         \toprule
         DPG & $\max{\sum_{p_i,p_j\in R}{\delta(\vec{pp_i}, \vec{pp_j})}}$ & Serial \\
         NSSG & $\delta(\vec{pp^{*}}, \vec{pp^{\prime}})>\gamma$ & Serial \\
         NSG & $dis(p, p^{\prime}) < dis(p^*,p^{\prime})$ & Serial \\
         Vamana & $dis(p, p^{\prime}) < \alpha *dis(p^*,p^{\prime})$ & Serial \\
         CAGRA & $\max(dis(p_i,p_k), dis(p_k,p_j)) < dis(p_i,p_j)$ & Unbalanced parallel\\
        \bottomrule
    \end{tabular}
    \label{tab:prun_summary}
\end{table}

\subsubsection{\textbf{Pruning Strategies.}}
\label{subsubsection:pruning}

Pruning strategies refine neighbor distributions in initialized $k$-NN graphs to improve query performance. Below are some notable approaches. 

DPG~\cite{li2019approximate} adjusts the neighbor distribution by maximizing the average angle between neighbors. Formally, for each node $p$ and its final neighbor list $R$, the objective is to maximize the angular dispersion $\max{\sum_{p_i,p_j\in R}{\delta(\vec{pp_i}, \vec{pp_j})}}$, where $\delta(\vec{a},\vec{b})$ denotes the angle between vectors $\vec{a}$ and $\vec{b}$. A greedy algorithm is designed to iteratively insert neighbors into $R$, maximizing this angular dispersion. 

NSSG~\cite{fu2021high} also considers the angle between neighbors. It sets a threshold $\gamma$. For each node $p$, the neighbors and 2-hop neighbors are collected into a candidate set $V$. The algorithm iteratively selects the nearest neighbor $p^{\prime}$ for $p$ from $V$. 
$p^{\prime}$ is inserted into $R$ if and only if $\forall p^{*}\in R$, $\delta(\vec{pp^{*}}, \vec{pp^{\prime}})>\gamma$.

NSG~\cite{fu2017nsg} claims that not all neighbors of a node contribute equally to search efficiency. For each node $p$, it first collects the candidate set $V$ by recording all visited nodes while searching $p$ on the \textit{k}-NN graph. It then iteratively selects the nearest node $p^{\prime}$ for $p$ from $V$. 
$p^{\prime}$ is inserted into $R$ iff $dis(p, p^{\prime}) < min_{p^*\in R}dis(p^*,p^{\prime})$.

Vamana~\cite{jayaram2019diskann} modifies NSG’s strategy by relaxing the proximity constraint to $dis(p, p^{\prime}) < \alpha *dis(p^*,p^{\prime})$, where $\alpha >1$, promoting long-range connections to improve graph navigability. 

CAGRA~\cite{ootomo2024cagra}, inspired by NGT~\cite{ngt2018}, takes an initialized $k$-NN graph as input, where it only stores the neighbor lists for all the nodes, without storing the exact distances between nodes and their neighbors. Instead, CAGRA estimates the distances between the node and its neighbors based on the position of the neighbors in the sorted neighbor list. For example, for a neighbor $X$ that is located at the first position in node $A$'s neighbor list, its position 1 is used as an estimate of $dis(A, X)$. 
CAGRA prunes edges based on a concept of a \textit{detourable route}. For an edge $(p_i, p_j)$, where $p_i$ and $p_j$ are nodes in the initialized graph, the edge has a detourable route $(p_i\rightarrow p_k\rightarrow p_j)$ if $\max(dis(p_i,p_k), dis(p_k,p_j)) < dis(p_i,p_j)$. This strategy prunes neighbors with more detourable routes. 

These pruning strategies are summarized in Table~\ref{tab:prun_summary}. Notably, DPG, NSSG, NSG, and Vamana rely on serial neighbor insertion to manage complex dependencies in neighbor lists. This process requires sequential validation of each candidate against existing neighbors before insertion, which inherently limits GPU parallelism due to dependency chains. While CAGRA’s pruning strategy enables parallel computation, it introduces unbalanced workloads among GPU threads due to the variable number of nodes accessed across neighbors, which further reduces GPU efficiency.

\subsection{GPU Architecture}

The memory architecture of the GPU comprises three primary tiers: global memory, shared memory, and registers. Global memory provides a large storage capacity but has comparatively lower access speeds. Shared memory, on the other hand, allows high-bandwidth data exchange within thread blocks, facilitating faster communication between threads.
Registers offer the fastest access latency, storing thread-specific data and ensuring isolation for each thread once allocated~\cite{cao2023gpu, owens2008gpu}.
In modern GPU architectures, there are two main types of computing units: CUDA cores and Tensor cores. CUDA cores function as general-purpose processors and are responsible for handling a wide range of computational workloads. Tensor cores, however, are specialized hardware designed for accelerating matrix operations, 
performing fixed-size matrix multiplications with high efficiency per clock cycle~\cite{zhu2019sparse}. 

\subsection{Overview of {\sf Tagore}}

\begin{figure}[tbp]
    \centering
    \vspace{-2mm}
    \includegraphics[width=1\linewidth]{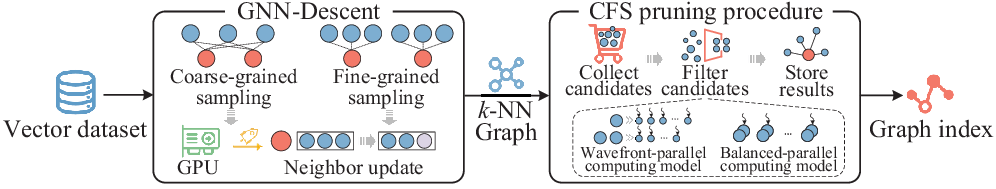}
    \vspace{-7mm}
    \caption{Index construction workflow of {\sf Tagore} when the dataset fits in GPU memory. }
    \vspace{-3mm}
    \label{fig:overview}
\end{figure}

{Figure \ref{fig:overview} illustrates the index construction pipeline in {\sf Tagore} for dataset fits within GPU memory. The framework first employs GNN-Descent, a GPU-accelerated two-phase $k$-NN graph initialization algorithm, to efficiently construct an approximate $k$-NN graph. Subsequently, the $k$-NN graph undergoes pruning via a CFS pruning framework, which unifies multiple pruning strategies while optimizing GPU utilization. For datasets exceeding GPU memory capacity, {\sf Tagore} automatically transitions to the GPU-CPU-disk asynchronous construction mode (Figure \ref{fig:asyn_framework}). During the asynchronous construction, the dataset is divided into clusters, with local indexes constructed on GPUs, and merged to disk using a cluster-aware caching mechanism to minimize I/O times. }
\section{Two-phase GNN-Descent}
\label{sec:gnn_descent}
As Tagore famously said, \textit{``The butterfly counts not months but moments''}, emphasizing the importance of valuing time. Similarly, an effective graph initialization algorithm embodies this principle by prioritizing speed, making the most of each moment in its execution. In this section, we explore the design of our novel GNN-Descent (GPU-based NN-Descent) algorithm integrated into {\sf Tagore}, demonstrating how it accelerates the initialization of \textit{k}-NN graphs. 

\subsection{Algorithmic Design}
\begin{figure}
    \centering
    \includegraphics[width=0.48\textwidth]{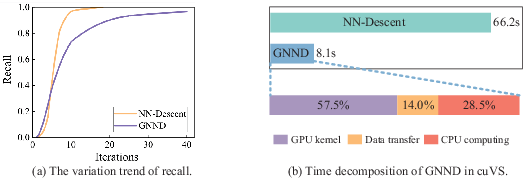}
    \vspace{-6mm}
    \caption{Recall of \textit{k}-nearest neighbors and time components for NN-Descent and GNND, where $k=96$ and dataset is SIFT. Figure \ref{fig:descent_motivation}(b) depicts the overhead to reach the recall of 95\%.}
    \label{fig:descent_motivation}
\end{figure}

\begin{figure}
    \centering
    \includegraphics[width=0.48\textwidth]{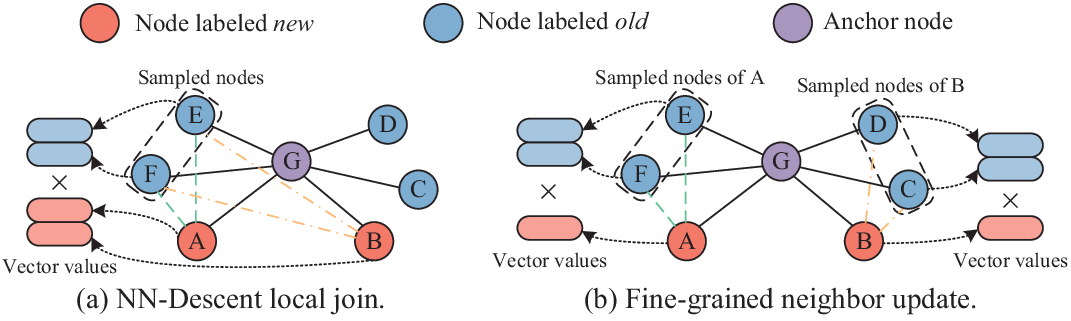}
    \vspace{-6mm}
    \caption{Analysis of NN-Descent sample strategy.}
    \label{fig:descent_problem}
    \vspace{-4mm}
\end{figure}

Figure~\ref{fig:descent_motivation}(a) shows the progression of the recall value for the $k$-NN (with $k=96$) on graphs constructed using the NN-Descent implemented in KGraph~\cite{kgraph2011} on CPU and GNND implemented in cuVS~\cite{wang2021fast,cuvs2024} on GPU. The dataset used is SIFT, which consists of 1 million vectors, each with 128 dimensions.
GNND exhibits a markedly slower recall improvement rate than the CPU-based NN-Descent under the same iterations, with convergence deceleration becoming more pronounced once recall exceeds 0.8. This slowdown is primarily due to three factors. First, GNND causes multiple nodes to share the same batch of sampling nodes during each iteration, hindering fine-grained neighbor updates. Second, limited GPU memory capacity restricts the number of \textit{new} and \textit{old} nodes processed in each iteration. Lastly, GNND updates the neighbor lists by selecting only the closest neighbor for each node per iteration, leading to inefficient use of intermediate computational results (see Section \ref{subsubsection:nn_descent}). 

To illustrate the first problem, consider the example subgraph in Figure~\ref{fig:descent_problem}, where anchor node $G$ mediates neighbor discovery among its connected nodes. In this subgraph, nodes $A$ and $B$ are labeled as \textit{new}, while $C,D,E$ and $F$ are labeled as \textit{old}. Due to GPU memory constraints, \textit{new} and \textit{old} nodes are sampled for \textit{local join}. Suppose the sample size is 2. In Figure \ref{fig:descent_problem}(a), $\{A,B\}$ and $\{E,F\}$ are sampled, requiring pairwise distance calculations between these sets to assess potential neighbor list updates. This sampling strategy causes node $A$ and $B$ to share the same samples, but as shown in Figure \ref{fig:descent_problem}(a), the true nearest neighbors of $B$ are $\{C,D\}$ not $\{E,F\}$. Consequently, the process of \textit{local join} fails to explore the actual nearest neighbors of certain nodes, leading to a severe slowdown in convergence after recall surpasses 0.8. Early iterations remain unaffected by this issue because the graph quality is low, and even suboptimal sampling yields incremental recall gains. The CPU-based NN-Descent avoids this limitation, as it can sample more nodes without memory restrictions, covering the nearest neighbors of more nodes~\cite{dong2011efficient}, but at the cost of computational redundancy from extraneous distance evaluations. 

\begin{figure*}[htbp]
    \centering
    \includegraphics[width=0.98\textwidth]{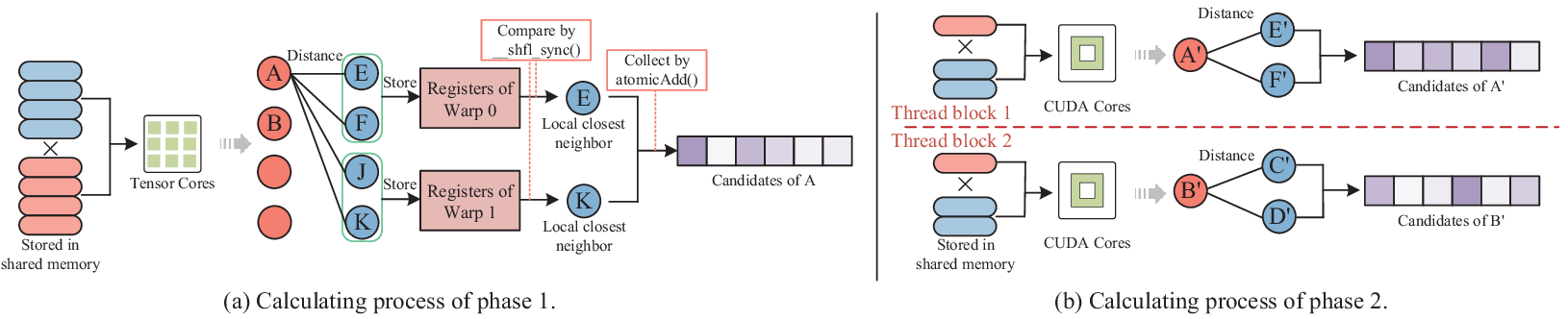}
    \vspace{-3mm}
    \caption{Distance Calculation process of GNN-Descent on GPU.}
    \label{fig:descent_calculating}
\end{figure*}

Given GPU memory constraints, it is infeasible to expand the sampling size as freely as on CPUs. 
To address this, we propose a fine-grained sampling strategy, shown in Figure \ref{fig:descent_problem}(b), where each node $v$ in the graph independently samples from the neighbor list of its top-$m$ nearest neighbors (with $m=2$). This ensures that the sampled nodes remain close to $v$, resolving the problem caused by shared sampling. However, this approach introduces a new problem: while the original \textit{local join} process leverages efficient GPU-optimized matrix multiplication for \textit{new-old} and \textit{new-new} pair-wise distance calculations, the independent sampling strategy in Figure \ref{fig:descent_problem}(b) fragments these operations into vector-matrix multiplication operations, reducing the computational throughput of GPUs. 

\setlength{\textfloatsep}{0pt}
\begin{algorithm}[tbp]
    \caption{Two-phase GNN-Descent}
    \label{alg:two_phase_descent}
    \LinesNumbered
    \KwIn{dataset $V$, numbers of iterations $it_1$ and $it_2$ for the first and second phases respective, degree $k$}
    \KwOut{\textit{k}-NN graph $G$}
    \For{$v\in V$ in parallel}{
        Initialize neighbor list $G[v]$ randomly; \\
    }
   \tcc{First phase}
    \For{($curIt \leftarrow 0$; $curIt < it_1$; $curIt++$)}{
        \For{$v \in V$ in parallel}{
            Sample nodes using shared sampling strategy;\\
            Perform \textit{local join} using matrix multiplication;\\
            Update closer neighbors to the corresponding neighbor lists;\\
        }
    }
    \tcc{Second phase}
    \For{($curIt\leftarrow 0$; $curIt < it_2$; $curIt++$)}
    {
        \For{$v \in V$ in parallel}{
            Sample neighbors of top-$m$ unvisited NN of $v$;\\
            Calculate distance between $v$ and sampled nodes;\\
            Update closer neighbors to $G[v]$;\\
        }
    }
    \Return{G}
\end{algorithm}
\setlength{\textfloatsep}{12pt plus 2pt minus 2pt}

To mitigate the limitations of shared sampling while maintaining efficiency, we propose a two-phase GNN-Descent algorithm tailored for GPUs, as outlined in Algorithm~\ref{alg:two_phase_descent}. 
The shared sampling strategy does not experience significant convergence deceleration during the early iterations, as the initial graph quality is low. Therefore, in the first phase, we retain the original NN-Descent sample strategy (Figure \ref{fig:descent_problem}(a)) to exploit its rapid convergence in the early stages (lines 4-7). 

After completing a total of $it_1$ iterations in the first phase, it switches to the second phase (lines 8-12). In the second phase, each node independently samples from the neighbor lists of its $m$ nearest neighbors, as shown in Figure \ref{fig:descent_problem}(b), ensuring fine-grained neighbor updates. Unlike the basic NN-Descent algorithm in \cite{dong2011efficient}, our approach enforces a non-revisitation policy, where nodes processed in previous iterations are excluded from subsequent sampling, thereby eliminating redundant computations.

\subsection{GPU Implementation}
Building on the GNN-Descent framework, we now turn to GPU-specific optimizations for each algorithmic component. In cuVS~\cite{cuvs2024}, NN-Descent is executed collaboratively across CPU and GPU: the GPU handles \textit{local join} operations, while the CPU manages neighbor list updates. Although they are pipelined, inefficiencies arise from frequent CPU-GPU data transfers, synchronization overhead, and the limited computational capacity of the CPU. For example, Figure \ref{fig:descent_motivation}(b) shows the time required to generate a graph that achieves a recall of 95\% in $k$-NN searches. While GNND significantly reduces the time compared to NN-Descent on CPU, there is still room for further improvement. The inefficiencies mentioned above collectively account for over 40\% of the total execution time of GNND, underscoring the need for GPU-native optimizations. To address these bottlenecks, we fully offload NN-Descent execution to the GPU and redesign key computational modules to optimize memory access patterns and parallel efficiency. Both phases of GNN-Descent include three operations: sampling, distance calculation, and neighbor list updates. While parallel sampling is straightforward, we 
focus on GPU-specific optimizations for distance computation and neighbor updates, which dominate runtime and memory overhead. 

\begin{figure}
    \centering
    \includegraphics[width=0.43\textwidth]{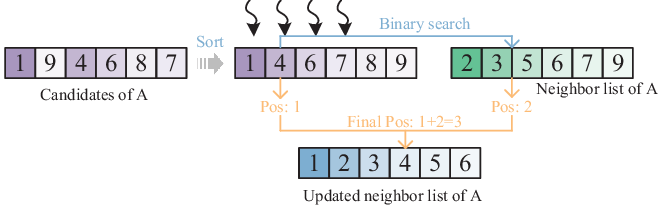}
    \vspace{-3mm}
    \caption{Parallel update process of GNN-Descent on GPU.}
    \label{fig:descent_update}
\end{figure}

Figure~\ref{fig:descent_calculating}(a) illustrates the first-phase distance calculation workflow, extending the example from Figure~\ref{fig:descent_problem} with an increased sample size of 4 to clarify the algorithm's mechanics. 
Similar to GNND, GNN-Descent initially loads the corresponding nodes into the shared memory and performs \textit{local join} via matrix multiplication on Tensor Cores. 
After the \textit{local join}, GNND retains only the closest neighbor for each node, 
wasting extensive computational results and exacerbating the convergence deceleration. To address this, GNN-Descent implements a data-reuse-aware strategy: after matrix multiplication in Tensor Cores, partial distance matrices are stored in the registers of each GPU warp (e.g., Warp 0 stores the distance between $A$ and $\{E,F\}$, while Warp 1 stores the distance between $A$ and $\{J,K\}$ in Figure \ref{fig:descent_calculating}(a)). 
We then select the closest node within each warp for each node (e.g., node $E$ for $A$ in Warp 0, and node $K$ for $A$ in Warp 1) and store them in the corresponding candidate list. The selection of the local closest node can be efficiently performed using the warp-level intrinsic \textit{\_\_shfl\_sync()} in CUDA for register-based value comparisons, avoiding memory access and enabling efficient multi-candidate retention. 

In the second phase, as shown in Figure \ref{fig:descent_calculating}(b), we implement node-centric parallelism by distributing the calculation of distances between each node and its neighbors in the candidate list
to individual thread blocks in the GPU. Within each block, the vector values of sampled nodes are loaded into shared memory, and vector-matrix multiplication is executed via CUDA Cores. Then, thread blocks compute distances between $A^\prime$ and the sampled candidates $\{E^\prime,F^\prime\}$, where the notation $X'$ refers to the vector of the node $X$.
To minimize computational overhead in subsequent updates, we retain only candidates closer than the farthest node in the current neighbor list while each list captures up to $k$ neighbors. 
This threshold guarantees that non-viable candidates are excluded early, as they cannot improve the neighbor list during the update phase.

To eliminate locking overhead and memory access during neighbor list updates, we propose a lock-free merging mechanism that uses binary search for deterministic ranking. As depicted in Figure~\ref{fig:descent_update}, the update module also uses node-centric parallelism. Within each thread block, the candidates for each node are sorted by their distance in shared memory. Subsequently, a thread is assigned to each element in both the candidate list and the neighbor list. These threads calculate the relative rank of their assigned element within the opponent's list via binary search. This rank, combined with the position of the element in its own sorted list, defines its insertion index in the merged neighbor list.

\section{Pruning Strategies}
\label{sec:pruning}

Tagore once said, \textit{``The best does not come alone. It comes with the company of all''}, reflecting a principle of collaborative optimization. 
In a similar spirit, this section introduces a generalized pipeline to accelerate mainstream refinement-based pruning strategies, unifying their execution into a modular GPU-optimized pipeline. Furthermore, we detail two specialized GPU kernels developed within the framework to achieve scalable, hardware-aware pruning.

\subsection{Analysis of Computing Procedure}
\label{subsec:unified_api}

Table \ref{tab:prun_summary} summarizes various mainstream pruning strategies used in existing refinement-based methods. Rather than accelerating individual strategies, our goal is to design a unified computing framework that can accelerate all such strategies. To achieve this, we first conduct a methodical analysis of the shared computational patterns across these strategies. We observe that all pruning strategies in Table \ref{tab:prun_summary} follow a three-step workflow: (1) \textit{Collect} candidate neighbors, (2) \textit{Filter} based on pruning strategies, and (3) \textit{Store} the final neighbors. We formalize this workflow as the CFS (Collect-Filter-Store) framework, as outlined in Algorithm~\ref{alg:pruning}. CFS preserves the node-centric parallelism model on GPUs, where each thread block is assigned to prune a single node’s candidate set, ensuring both scalability and hardware efficiency. 

\setlength{\textfloatsep}{0pt}
\begin{algorithm}[tp]
    \caption{The CFS procedure of pruning}
    \label{alg:pruning}
    \LinesNumbered
    \KwIn{dataset $V$, \textit{k}-NN graph $G$}
    \KwOut{final graph index $G'$}
    \For{$v\in V$ in parallel}{
        \tcc{The built-in options of parameter Mode include `1-hop', `2-hop', and `path'}
        Fill candidate set $Cand$ by {\tt Collect}(Mode, CandSize, $v$);\\
        \tcc{The built-in options of parameter Metric include `dist', `angle', and `rank'}
        Filter the candidates with {\tt Filter}(Metric, Thres, $Cand$);\\
        Store results into the index using {\tt Store}($G'[v]$, $Cand$);\\
    } 
\end{algorithm}
\setlength{\textfloatsep}{12pt plus 2pt minus 2pt}

\noindent\textbf{Collect.} The {\tt Collect} operation identifies candidate neighbors for each node from the \textit{k}-NN graph. The pruning strategies listed in Table \ref{tab:prun_summary} employ various methods to collect candidates. For instance, DPG and CAGRA limit candidate selection to immediate neighbors (1-hop) in the \textit{k}-NN graph, while NSSG expands to 2-hop neighbors in the \textit{k}-NN graph for broader exploration. In contrast, NSG and Vamana initiate searches from the entry point in the \textit{k}-NN graph, collecting all visited nodes along the search path as candidates, which are then filtered in the next phase. To unify these approaches, the {\tt Collect} operation exposes a configurable parameter $Mode$ (line 2 of Algorithm~\ref{alg:pruning}), which supports three options: `1-hop', `2-hop' and `path'. Collecting 1-hop and 2-hop neighbors in parallel on GPUs is straightforward, while the `path' option leverages CAGRA's batched search method~\cite{ootomo2024cagra} to generate candidates with low latency. 

\noindent\textbf{Filter.} The {\tt Filter} operation is computationally dominant during pruning. As shown in line 3 of Algorithm~\ref{alg:pruning}, it filters the candidates in $Cand$ based on a user-specified pruning strategy. 
As listed in Table \ref{tab:prun_summary}, refinement-based pruning strategies fall into three categories: distance-based, angle-based, and rank-based. NSG and Vamana are distance-based strategies that prune candidates according to the distance between the corresponding nodes. DPG and NSSG are angle-based strategies, while CAGRA adopts a rank-based strategy. To accommodate these variations, {\tt Filter} exposes a parameter $Metric$ to allow users to select their desired category. Moreover, the parameter $Thres$ controls the threshold for pruning strategies such as $\alpha$ in Vamana and $\gamma$ in NSSG. When $\alpha=1$, Vamana degenerates to NSG, and setting $\gamma=0$ invokes DPG. Given its computational intensity and the complex neighbor dependencies, Section \ref{subsec:opt_GPU} elaborates on our GPU-specific optimizations for {\tt Filter}. 

\noindent\textbf{Store.} To optimize memory access, candidate nodes are temporarily stored in high-speed shared memory in the {\tt Filter} operation. The {\tt Store} operation then writes the final set of neighbors from shared memory to their pre-allocated positions within GPU global memory, and then to the CPU for storage in files.

{The CFS framework provides a generalized abstraction for diverse pruning strategies, as summarized in Table \ref{tab:prun_summary}. Notably, it accommodates a broad spectrum of pruning strategies with minimal modifications\footnote[2]{Examples are available at https://github.com/ZJU-DAILY/Tagore}, since the methods in Table \ref{tab:prun_summary} represent fundamental strategies that influence numerous derivative strategies. Importantly, the framework maintains extensibility for future pruning paradigms that may diverge from the current methodologies. Users can selectively override specific functions while preserving the overall pipeline architecture, thereby ensuring both backward compatibility and forward flexibility. }

\subsection{Computing Optimizations on GPUs}
\label{subsec:opt_GPU}

Within the CFS framework introduced in Section \ref{subsec:unified_api}, {\tt Filter} is the most computationally intensive operation due to its central role in pruning computations. Therefore, this subsection focuses on the computational paradigm underlying {\tt Filter} across pruning strategies and tailors the paradigm for GPU-specific optimizations. 

\begin{figure}
    \centering
    \includegraphics[width=0.48\textwidth]{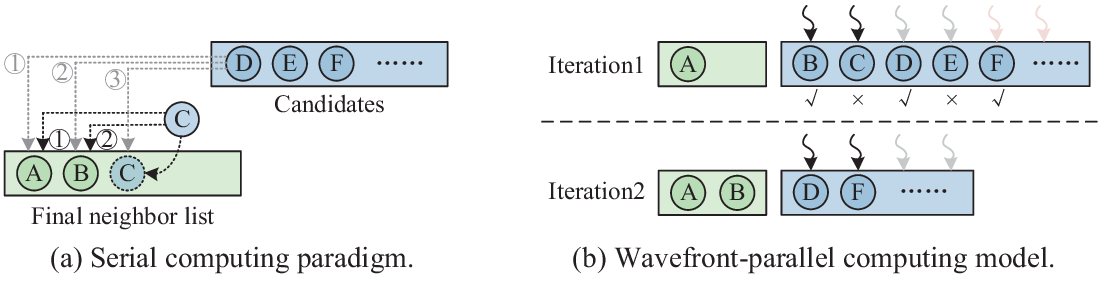}
    \vspace{-7mm}
    \caption{Serial computing paradigm of {\tt Filter}.}
    \label{fig:incrental}
\end{figure}

DPG, NSSG, NSG, and Vamana employ a serial computing paradigm characterized by sequential data dependencies. As depicted in Figure~\ref{fig:incrental}(a), consider nodes $A$ and $B$ already in the neighbor list. 
When node $C$ is retrieved from the candidate set, its eligibility for insertion depends on computing pairwise distances/angles with all existing nodes in the neighbor list (here, $A$ and $B$ in Figure \ref{fig:incrental}(a)), according to the pruning strategies in Table~\ref{tab:prun_summary}. 
Then, node $D$'s eligibility depends on $C$'s updated status, necessitating strict sequential processing. This dependency chain—where each candidate’s evaluation hinges on prior results—forces a sequential execution paradigm, which is inherently incompatible with GPU parallelism. 

To enable high-performance pruning for serial computation strategies, we propose a wavefront-parallel computing model specifically designed for GPUs to exploit their massive parallelism. As illustrated in Figure \ref{fig:incrental}(b), each thread block allocates several warps (e.g., 2 warps in our example) to process candidate nodes in parallel. In the following, we explain how to populate the neighbor list with nodes from the candidate pool.
Each warp independently evaluates the candidate node against the current neighbor list. For example, in iteration 1, the two warps calculate the distance or angle between $A$ and $B$, and that between $A$ and $C$, respectively. If the number of warps is less than the number of candidates, candidates are processed iteratively.
Candidates pruned by the pruning strategies are removed from the candidate set at the end of the iteration (e.g., nodes $C$ and $E$ are pruned away from the candidate set in iteration 1).
Subsequently, in iteration 2, an arbitrary candidate (e.g., $B$) within the current candidate set is inserted into the neighbor list.
The remaining candidates (i.e., $D$ and $F$) are re-evaluated in subsequent iterations,
and we only need to test if they conflict with the node newly added into the neighbor list (e.g., $B$ in iteration 2). This model transforms the inherently sequential dependencies into \textit{wavefront parallelism}: candidates are processed in batches, and each iteration resolves dependencies only within its wavefront. 

\begin{figure}
    \centering
    \includegraphics[width=0.46\textwidth]{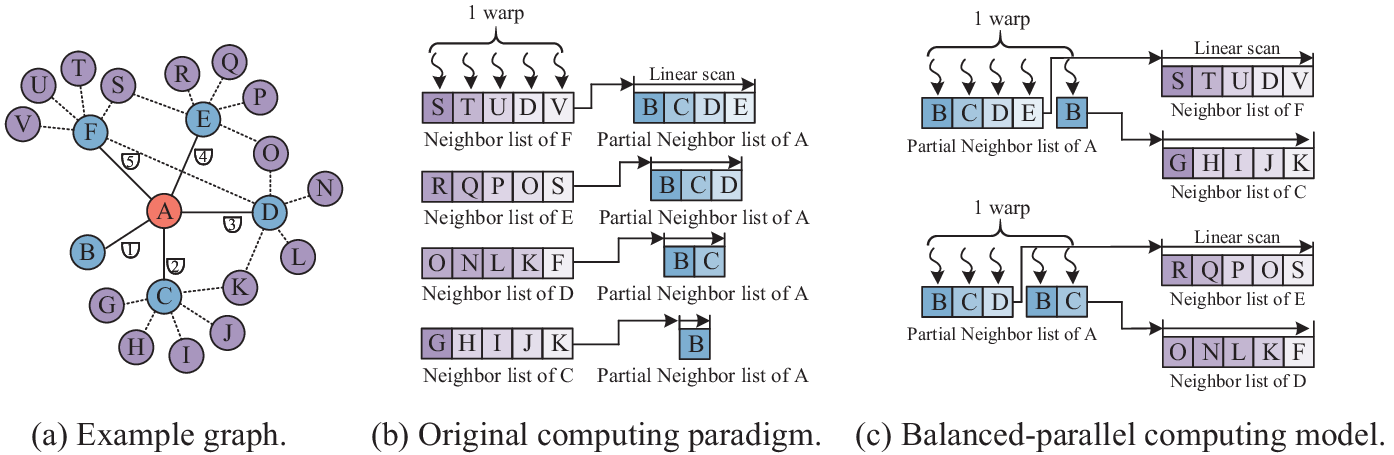}
    \vspace{-3mm}
    \caption{Unbalanced parallel paradigm of {\tt Filter}.}
    \label{fig:cagra}
    \vspace{-4mm}
\end{figure}

CAGRA adopts a distinct unbalanced parallel paradigm, differing from the serial paradigm. It utilizes the rank of node $A$'s neighbors (i.e., the position $pos_u$ of each neighbor $u$ in $A$'s neighbor list) as an estimate of the distance between $A$ and its neighbors (e.g., $dis(A,B)=pos_B=1$ and $dis(A,C)=pos_C=2$, as shown in Figure~\ref{fig:cagra}(a)). 
As depicted in Figure~\ref{fig:cagra}(b), for each candidate neighbor of $A$, CAGRA employs a warp of threads to retrieve the candidate's neighbors. Each thread then scans $A$'s neighbor list to identify shared neighbors. For instance, when a thread retrieves $D$ from the neighbor list of $F$ 
and detects $D$ is in $A$'s list, it identifies the path $A\rightarrow D \rightarrow F$. As $dis(F,D)=4<dis(A,F)=5$, the path $A\rightarrow D \rightarrow F$ is flagged as a detourable path for the edge $(A,F)$.
The candidate neighbors of $A$ are then pruned based on the number of detourable paths the corresponding edges have. 
However, this computational paradigm for detourable paths, shown in Figure \ref{fig:cagra}(b), cannot fully utilize GPU parallelism, as it employs only a single warp for calculation. 
Simply employing multiple warps to parallelize the calculation for each candidate leads to unbalanced workloads across warps, because different candidates (e.g., $F$, $E$, $D$, and $C$) need to scan varying numbers of $A$'s neighbors.
For example, node $A$ has five neighbors. Candidate $F$ needs to scan 4 of them, as neighbor $F$ can be skipped; while candidate $E$ needs to scan 3, as both $E$ and $F$ can be skipped. 
Note that $E$ is skipped because $A\rightarrow E\rightarrow E$ cannot be a detourable path of $A\rightarrow E$. Similarly, $F$ is skipped, as $dis(A,E)=4>dis(A,D)=3$ and $dis(A,F)=5>dis(A,D)$ and hence the path $A\rightarrow F\rightarrow E$ cannot be a detourable path either. 

To address the issue of unbalanced computing, we design a \textit{balanced-parallel computing model}, as shown in Figure \ref{fig:cagra}(c). It deploys two warps for each candidate pair $(u, v)$, where their positions satisfy $pos_u + pos_v=k+2$ (with $k$ denoting the degree of the \textit{k}-NN graph). 
This pairing ensures that the aggregate workload for each warp is balanced across warps. Each thread within a warp retrieves a node from the neighbor list of $A$ and scans the neighbor list of the corresponding candidate. Upon identifying a shared node that satisfies the detour path criteria, the thread atomically increments the detourable path count for the corresponding candidates. 

The two optimized GPU kernels enable users to simply modify the pruning equations for newly proposed pruning strategies with the serial and unbalanced parallel computing paradigms, without considering the complicated underlying implementation. 
\section{Large-scale Graph Indexing}
\label{sec:large_graph}

Tagore said, \textit{``The scabbard is content to be dull when it protects the keenness of the sword''}. Similarly, CPU and disk resources, much like the scabbard, play an essential, supporting role in protecting the efficiency of GPUs when constructing billion-scale graph indexes under stringent memory constraints.
In this section, we propose an asynchronous GPU-CPU-disk pipeline to efficiently build large-scale global graph indexes and introduce a cluster-aware cache mechanism to reduce I/O overhead between the CPU and disk. 

\subsection{GPU-CPU-disk Asynchronous Framework}
\label{subsec:asyn_framework}

Given the prohibitive cost and capacity limitations of GPU and CPU memory, disk storage provides a practical solution for large-scale indexing. Recent work, such as DiskANN~\cite{jayaram2019diskann}, SPANN~\cite{chen2021spann}, and Starling~\cite{wang2024starling}, explores disk-based indexing in scenarios with constrained CPU memory. 
However, to the best of our knowledge, no existing method efficiently constructs billion-scale global graph indexes when \textit{both GPU and CPU memory are constrained}.
To tackle this issue, in {\sf Tagore}, we adopt the workflow of DiskANN and extend it to a heterogeneous GPU-CPU-disk system. 

DiskANN partitions each node to its $m$ ($m>1$) closest clusters via \textit{k}-means. It then constructs local graph indexes for each cluster and writes these indexes to disk, finally merging the local indexes into a global index. 
However, directly mapping DiskANN's workflow to the GPU-CPU-disk heterogeneous system introduces two critical limitations. First, offloading local index construction to GPUs leaves CPUs idle, wasting their computational potential. Second, writing and reading local indexes to/from disk during merging incurs excessive I/O overhead, significantly reducing throughput. 

\begin{figure}
    \centering
    \includegraphics[width=0.46\textwidth]{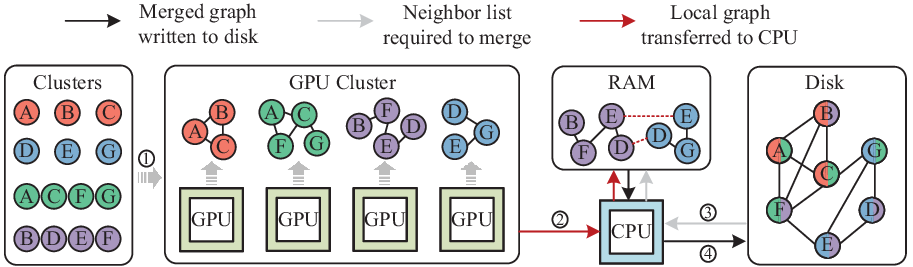}
    \vspace{-3mm}
    \caption{Asynchronous graph construction framework.}
    \label{fig:asyn_framework}
    \vspace{-4mm}
\end{figure}

\setlength{\textfloatsep}{0pt}
\begin{algorithm}[tp]
    \caption{The merging workflow of CPU.}
    \label{alg:CPU_merge}
    \LinesNumbered
    \KwIn{cluster set $CS$}
    \KwOut{global graph index $G$}
    Initialize $neiLoc[v]$ for each node $v$ in the dataset as $null$;\\ 
    \For{each cluster $C\in CS$}{
        Waiting for the GPU to transfer the local index $LI$;\\
        \For{each node $v\in C$}{
            \If{$neiLoc[v]=null$}{
            \tcp{Fetch the neighbor list pointer of $v$}
                $neiLoc[v]\leftarrow LI[v]$;\\
            }
            \Else{
            \tcp{Neighbor list of $v$ resides in disk}
                \If{$neiLoc[v]=-1$}{
                    Load neighbor list of $v$ into $neiLoc[v]$;\\
                }
                $neiLoc[v]\leftarrow merge(neiLoc[v],LI[v])$;\\
            }
        }
        \If{Cache in RAM is full}{
            \tcp{Evict a local index to disk}
            \For{each node $u$ in the evicted local index}{
                Write $neiLoc[u]$ to disk;\\
                $neiLoc[u]\leftarrow -1$;\\
            }
        }
    } 
\end{algorithm}
\setlength{\textfloatsep}{12pt plus 2pt minus 2pt}

To address these challenges, we propose an asynchronous construction framework that integrates the CPU into the pipeline and repurposes RAM as a cache to reduce I/O bottlenecks. 
As illustrated in Figure~\ref{fig:asyn_framework}, after distributing the data into clusters via \textit{k}-means as DiskANN does, {\sf Tagore} streams these clusters to GPUs for distributed local index construction. 
It supports multi-GPU parallelism, enabling simultaneous construction across multiple GPUs. Each GPU is paired with a dedicated CPU thread that orchestrates data transfers and execution control. Upon completing a local index, the CPU thread writes the index to RAM while transferring the next cluster to the GPU, ensuring continuous GPU utilization.

Concurrently, the CPU initiates the merging process for local indexes stored in RAM, as formalized in Algorithm~\ref{alg:CPU_merge}. {\sf Tagore} maintains an array $neiLoc$ to track the memory offsets of all nodes' neighbor lists. The CPU merges local indexes by scanning each node $v$ in the cluster (line 4). When a node is first visited, its neighbor list location is recorded in $neiLoc$ (lines 5-6). For previously visited nodes, the CPU verifies whether the neighbor list from the prior access has been persisted to disk (line 8). If yes (indicated by value -1 of $neiLoc[v]$), it fetches the list from disk (line 9). Otherwise, the neighbor list of $v$ must be cached. The CPU retrieves it and merges the existing neighbor list with the current one (line 10).  
When the cache reaches its full capacity, the CPU evicts one of the local indexes $LI$ in the cache into the disk, using the mechanism introduced in Section \ref{subsec:caching} (lines 11-14). 

The asynchronous graph construction framework establishes a heterogeneous computing pipeline where GPUs focus on parallel local index construction and CPUs concurrently merge partial indexes. It utilizes the computing resources of CPUs and employs RAM as a dynamic cache to reduce disk I/O overhead. 

\textbf{Discussion}. {While merging subgraphs on the GPU and streaming them directly to NVMe SSDs presents an appealing optimization, we avoid GPU-to-NVMe SSD data transfers to prioritize broad accessibility of our framework. The direct access of GPU-SSD is a feature limited to data-center-grade GPUs (e.g., NVIDIA A100/H100) and remains unsupported on widely used consumer-grade hardware (e.g., NVIDIA RTX 3090/4090), which would significantly limit practical adoption. 
Additionally, GPU-based subgraph merging necessitates retaining at least two subgraphs in GPU memory; otherwise, it would introduce additional data transfer from CPU or SSD to GPU. However, storing two subgraphs can further reduce the already limited GPU memory, which negatively impacts the efficiency of index construction.}

\subsection{Cluster-aware Caching Mechanism}
\label{subsec:caching}

\begin{figure}
    \centering
    \includegraphics[width=0.46\textwidth]{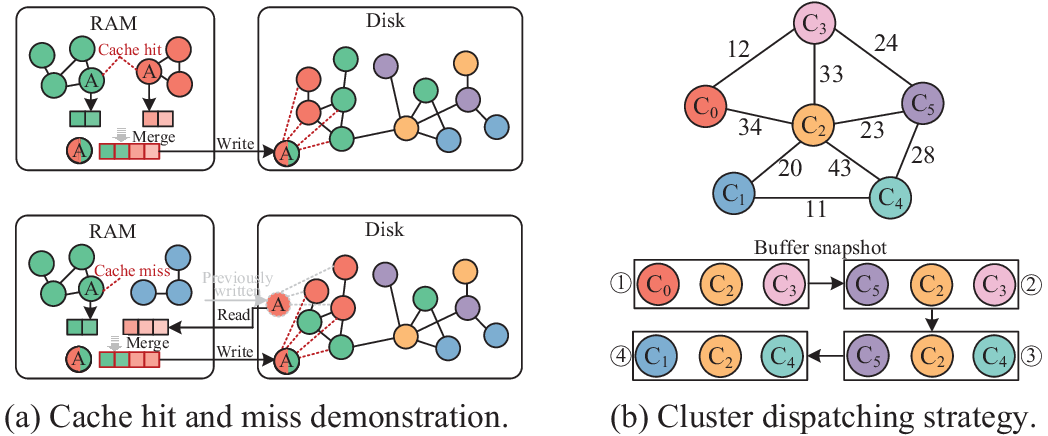}
    \vspace{-3mm}
    \caption{Cache mechanism based on cluster dispatching.}
    \label{fig:cache_mechanism}
\end{figure}

In Section \ref{subsec:asyn_framework}, we allocate a cache in RAM to reduce the I/O overhead between the CPU and disk. 
As depicted in Figure \ref{fig:cache_mechanism}(a), if the neighbor list of node $A$ resides in RAM, the cache is hit.
In this situation, the neighbor lists of the $A$ can be merged directly in RAM (line 10 in Algorithm \ref{alg:CPU_merge}). 
On the other hand, if the neighbor list of $A$ resides in disk (Figure \ref{fig:cache_mechanism}(b)), we have to load the neighbor list previously written from the disk (lines 8-9 in Algorithm \ref{alg:CPU_merge}), and then merge it with the current neighbor list (line 10 in Algorithm \ref{alg:CPU_merge}). 
As compared to the cache hit situation, the cache miss situation triggers the execution of lines 8-9 in Algorithm \ref{alg:CPU_merge}, which requires two extra I/O operations (the loading of $v$'s neighbor list from disk and previously written to disk).

The cache hit rate is intrinsically tied to the number of overlapping nodes retained in RAM. Thus, the cluster scheduling policy (determining the order in which clusters are loaded into or evicted from the cache) is critical for maximizing cache efficiency. 
To formalize this, as depicted in Figure \ref{fig:cache_mechanism}(b), we model clusters as vertices in an undirected graph $CG$, where an edge between clusters $C_i$ and $C_j$ exists if they share nodes, with edge weight $W(C_i, C_j)$ representing the number of shared nodes. 
Given a cache capacity of $n$ local indexes, our objective reduces to a graph traversal problem: selecting a sequence of vertices that maximizes cumulative edge weights (i.e., overlapping nodes) within a sliding window of $n$ vertices. This is equivalent to prioritizing vertices with high interconnectivity, ensuring maximal reuse of cached nodes during merging.

To achieve this goal, we traverse the graph using a greedy strategy that prioritizes clusters with the highest cumulative edge weights relative to those already in the cache. The pseudo-code for this process is presented in Algorithm~\ref{alg:cluster_order}. We begin by initializing the cache with a random cluster (e.g., the first cluster in our implementation), which has minimal impact on its performance (line 1). Next, the algorithm iteratively selects the cluster in $C$ that maximizes the sum of edge weights, until the cache reaches its full capacity (lines 2-4). When a cluster $C^{*}$ is identified, it is removed from the cluster set $C$ (which corresponds to the vertex set of the cluster graph), and a tuple $(C^{*},null)$ is appended to the dispatching order set $R$ (line 4).
Note that each tuple $(C_1, C_2)\in R$ denotes the loaded cluster $C_1$ and, if applicable, the evicted cluster $C_2$. 
Once the cache is full, loading additional clusters requires evicting existing ones. To determine which clusters in the cache to evict and which clusters in $C$ to load, we identify two clusters: $C^*\in C$ and $C'\in buf$, such that $\sum_{C_k\in (buf-\{C'\})}W(C^*, C_k)$ is maximized (line 7). 
This process of loading new clusters into the cache while evicting existing clusters continues until all the clusters are loaded. 

\setlength{\textfloatsep}{0pt}
\begin{algorithm}[tp]
    \caption{Cluster-aware dispatching strategy.}
    \label{alg:cluster_order}
    \LinesNumbered
    \KwIn{cluster graph $CG(C,E)$, cache size $n$}
    \KwOut{dispatching order $R$}
    \tcc{Initialize with the first cluster}
    $buf\leftarrow \{C[0]\}$; $C\gets C-\{C[0]\}$; $R\gets \emptyset$;\\
    \While{$|buf| < n$}{
        $C^{*}\leftarrow \underset{C_i\in C}{\operatorname{arg\,max}}\, \sum\nolimits_{C_j\in buf}W(C_i,C_j)$;\\
        $buf\leftarrow buf\cup\{C^{*}\}$;
        $C\leftarrow C-\{C^{*}\}$, $R\leftarrow R\cup\{(C^{*},null\}$;\\
    } 
    \While{$C$ is not empty}{
        $C^*,C^{\prime}\leftarrow \underset{C^*\in C,C^{\prime}\in buf}{\operatorname{arg\,max}}\, (\sum_{C_k\in (buf-\{C'\})}W(C^*, C_k))$;\\
        $buf\leftarrow buf-\{C^{\prime}\}$, $buf\leftarrow buf\cup\{C^{*}\}$;\\
        $C\leftarrow C-\{C^{*}\}$, $R\leftarrow R\cup\{(C^{*},C^{\prime})\}$;\\
    }
    \Return{R}
\end{algorithm}
\setlength{\textfloatsep}{12pt plus 2pt minus 2pt}

\noindent\textbf{Example.} Figure \ref{fig:cache_mechanism}(b) depicts a simple example of the dispatching strategy. The buffer size is 3, and there are a total of 6 clusters. The cache is initialized with $C_0$ and appended with ($C_2$, $null$) and ($C_3$, $null$) in turn.
Now the cache is full, and subsequent loading of local indexes requires eviction of existing ones. Next, $C_5\in C$ and $C_0\in buf$ are selected based on the criterion listed on line 6 of Algorithm~\ref{alg:cluster_order}. 
Accordingly, $C_0$ is evicted and $C_5$ is loaded as shown in buffer snapshot \ding{173}. Similarly, we get buffer snapshot \ding{174} and \ding{175}.   
\section{Experiments}
\label{sec:experiments}
In this section, we evaluate the performance of {\sf Tagore} and conduct comparative evaluations with existing graph indexing methods. 

\subsection{Experiment Settings}
\label{subsec:setting}

\noindent\textbf{Datasets.}
For comprehensive evaluations, we use 7 real-world datasets, including 5 million-scale and 2 billion-scale, which have been widely used in related works~\cite{gou2025symphonyqg,fu2017nsg,wang2021fast,ootomo2024cagra,jayaram2019diskann,cao2023gpu}. For the datasets that don't provide the query vectors, we randomly sample them from the data vectors. Table \ref{tab:dataset} summarizes the detailed properties of each dataset. 

\begin{figure*}
    \centering    
    \includegraphics[width=0.98\textwidth]{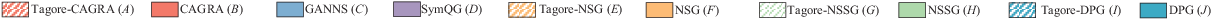}\\
    \vspace{-2mm}
    \hspace{-6mm}
    \subfigure[Deep-1M]{
    \includegraphics[width=0.2\textwidth]{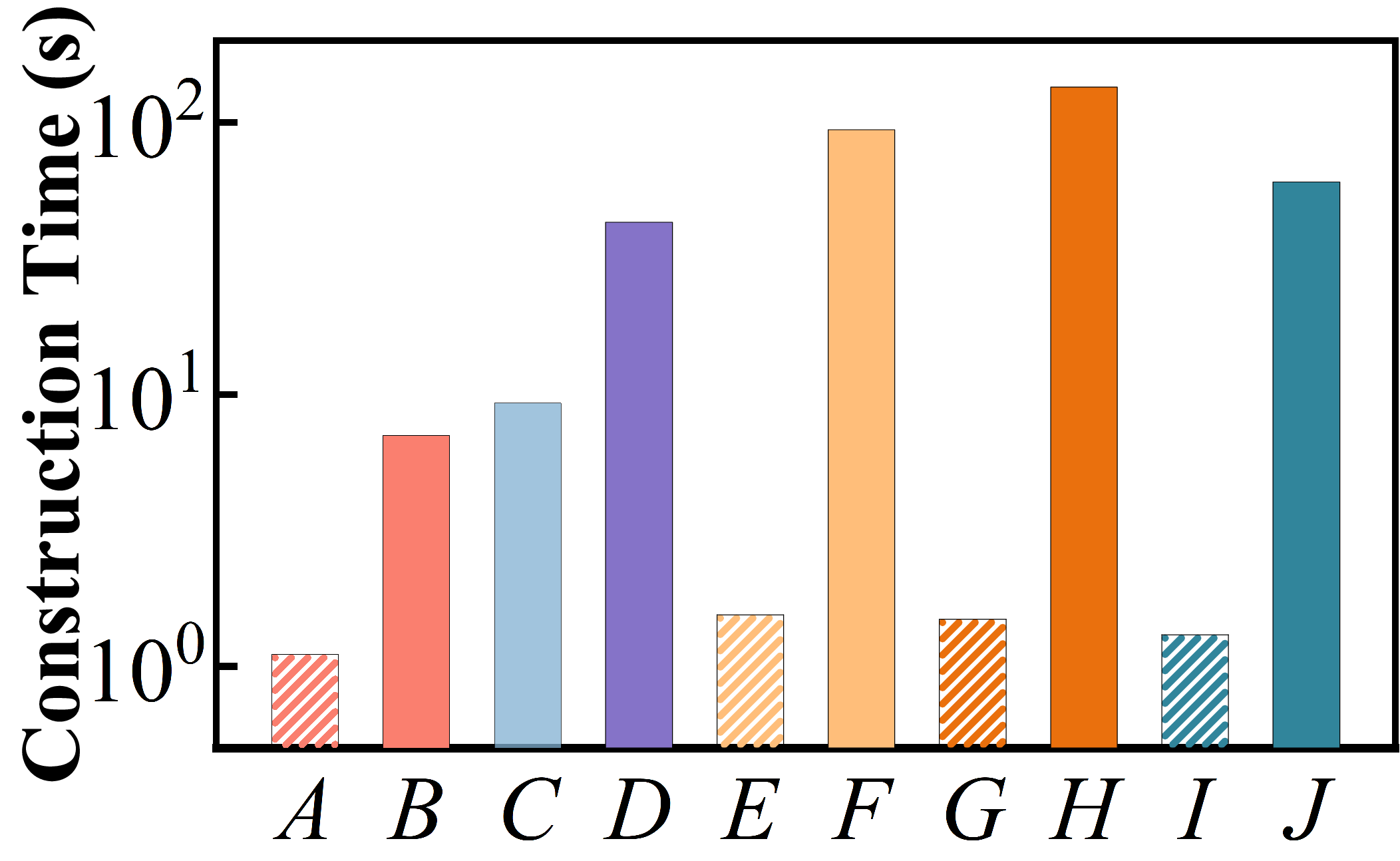}}
    \hspace{-2.2mm}
    \subfigure[SIFT]{
    \includegraphics[width=0.2\textwidth]{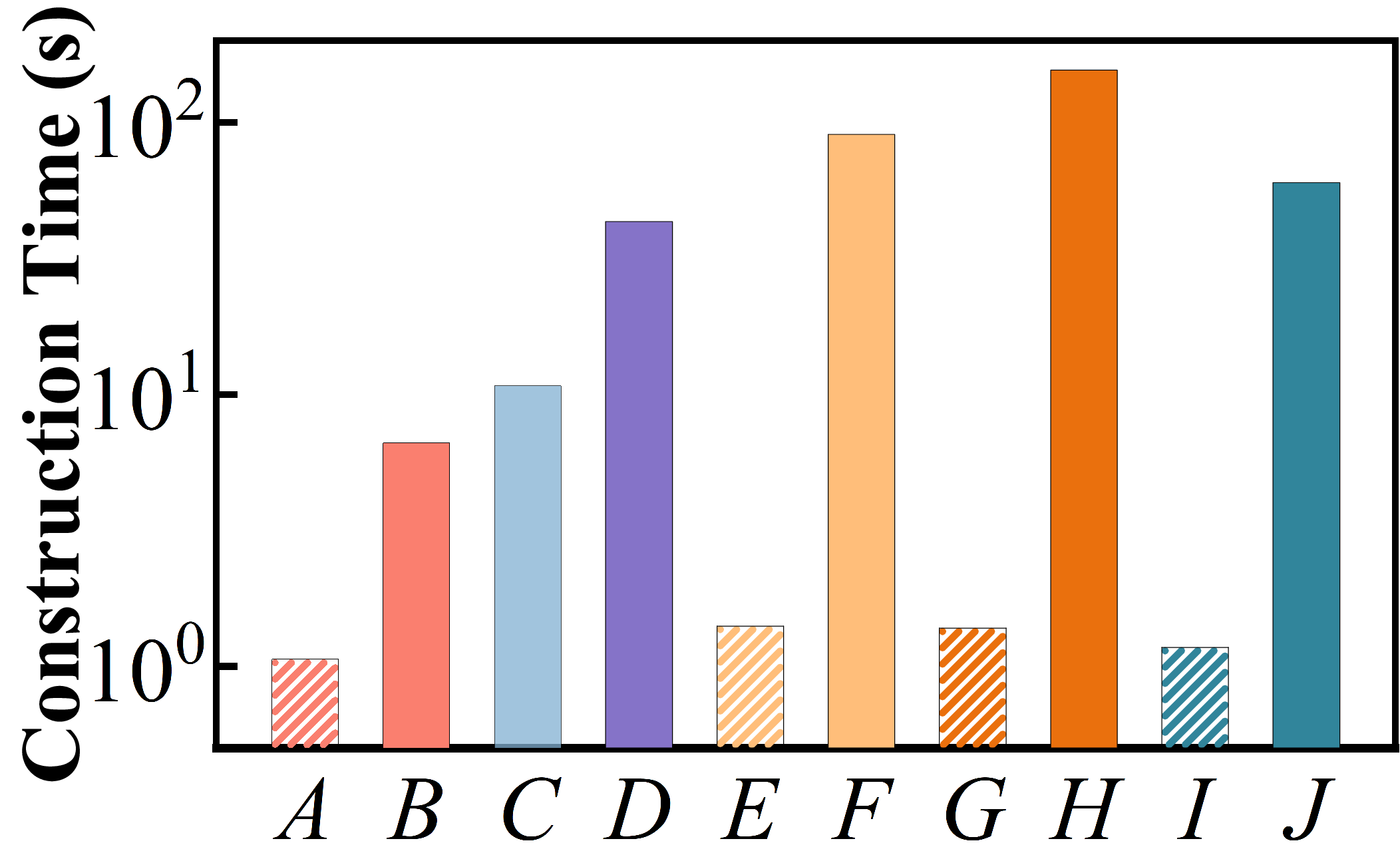}}
    \hspace{-2.2mm}
    \subfigure[UKBench]{
    \includegraphics[width=0.2\textwidth]{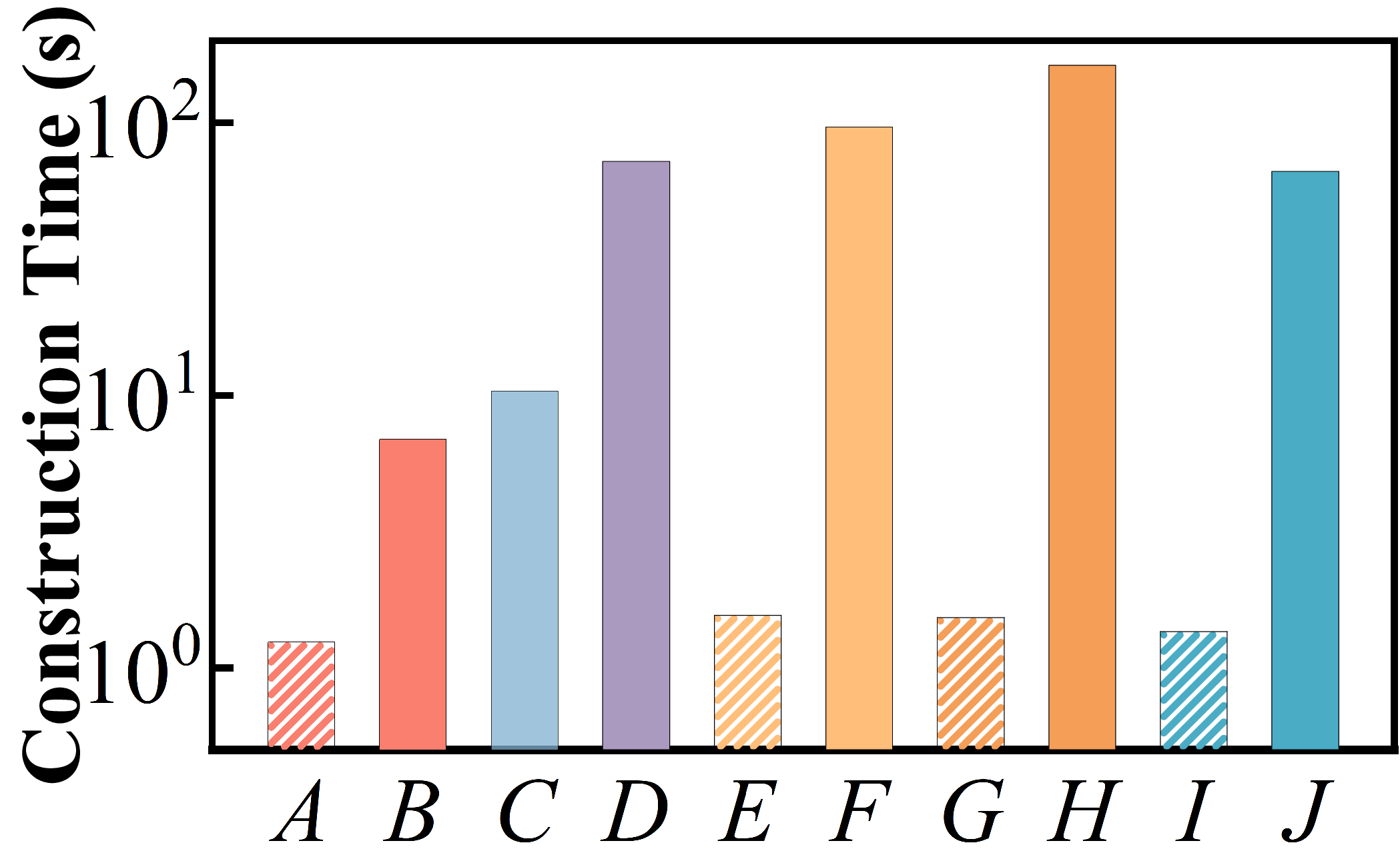}}
    \hspace{-2.2mm}
    \subfigure[Color]{
    \includegraphics[width=0.2\textwidth]{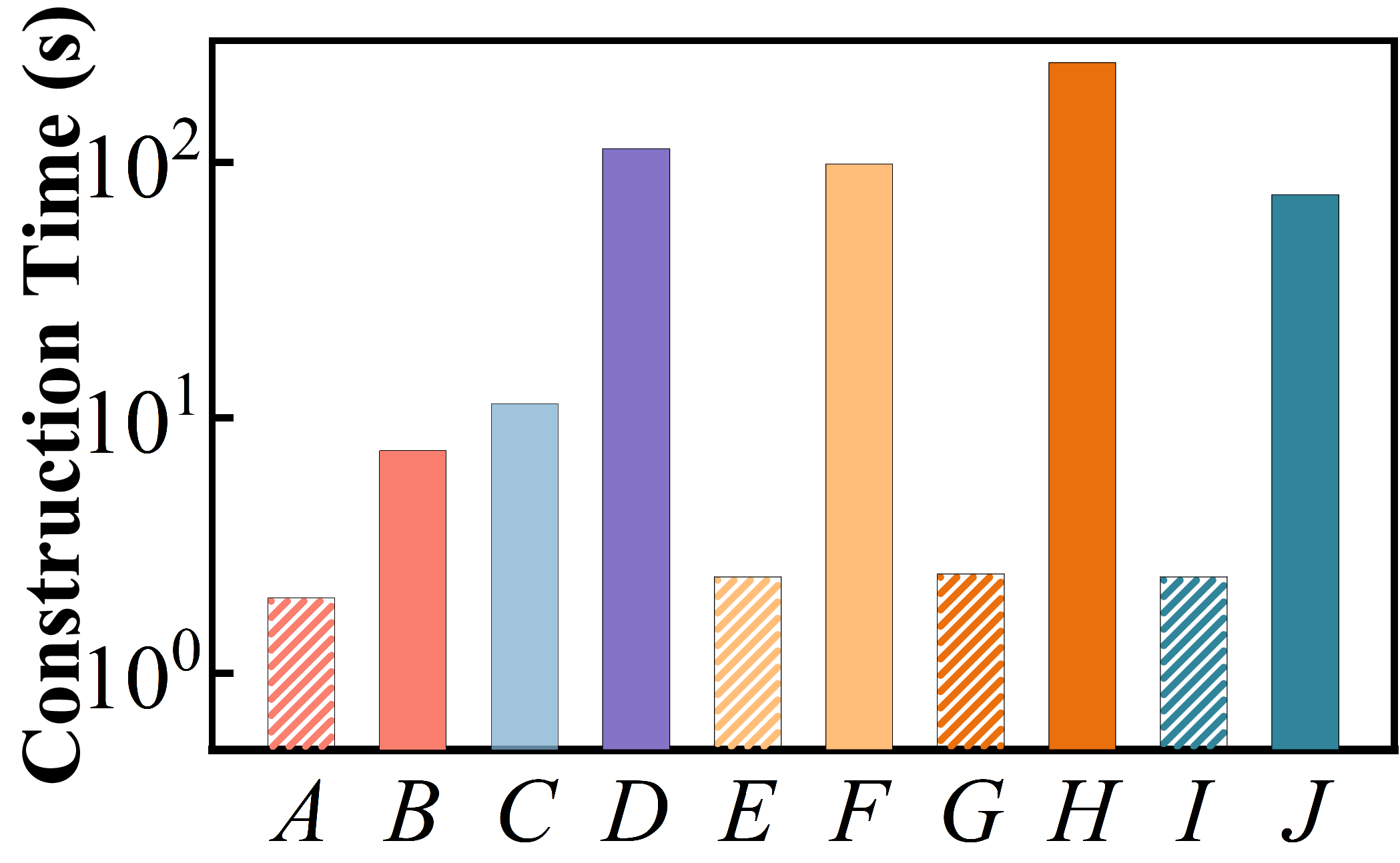}}
    \hspace{-2.2mm}
    \subfigure[Gist]{
    \includegraphics[width=0.2\textwidth]{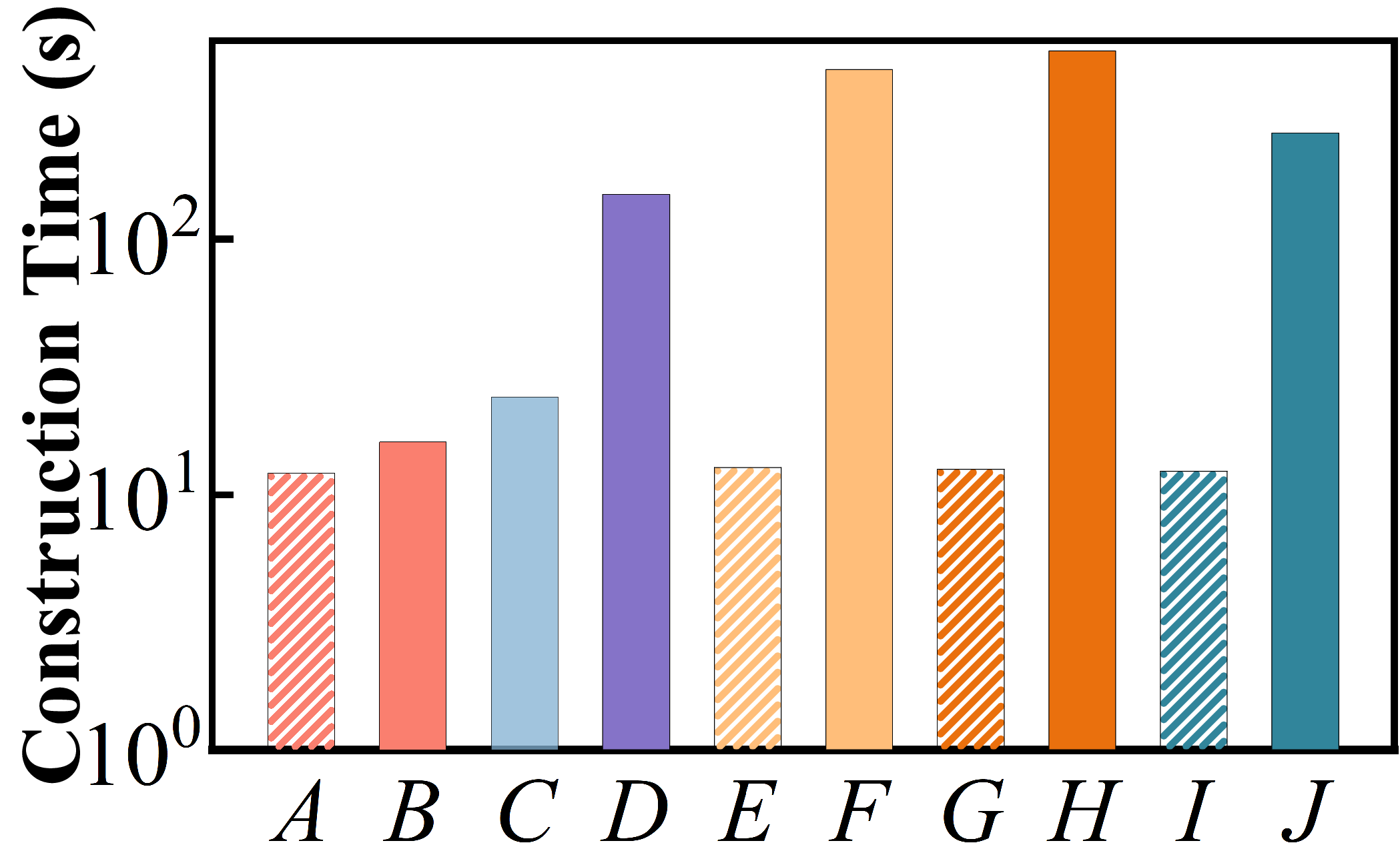}}
    \vspace{-5mm}
    \caption{Overall index construction time on million-scale datasets.
    }
    \label{fig:over_performance}
\end{figure*}

\begin{figure*}
    \centering    
     \vspace{-1mm}
    \includegraphics[width=0.98\textwidth]{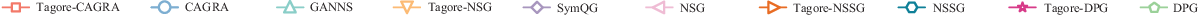}\\
    \vspace{-2mm}
    \hspace{-6mm}
    \subfigure[Deep-1M]{
    \includegraphics[width=0.2\textwidth]{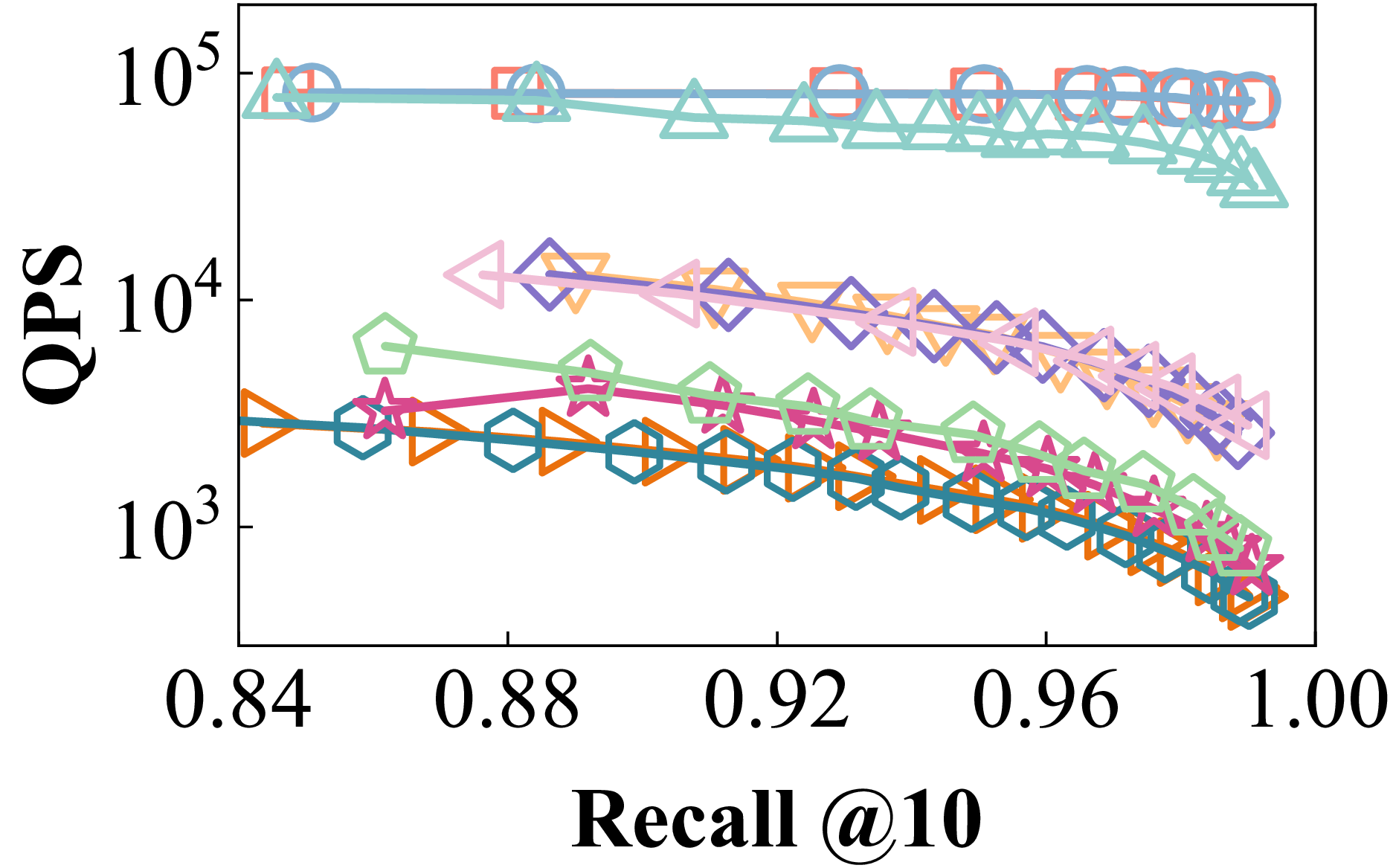}}
    \hspace{-2.2mm}
    \subfigure[SIFT]{
    \includegraphics[width=0.2\textwidth]{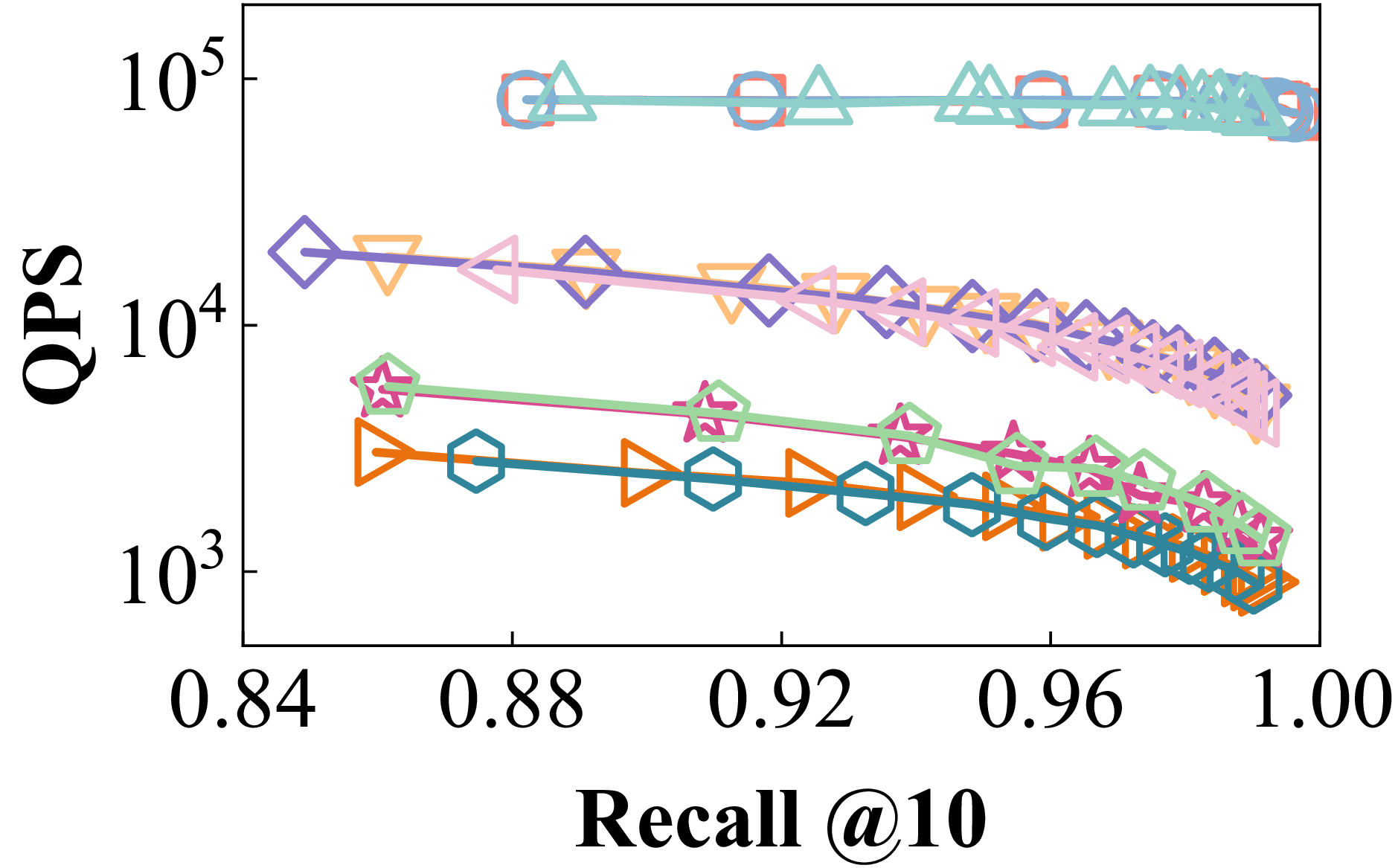}}
    \hspace{-2.2mm}
    \subfigure[UKBench]{
    \includegraphics[width=0.2\textwidth]{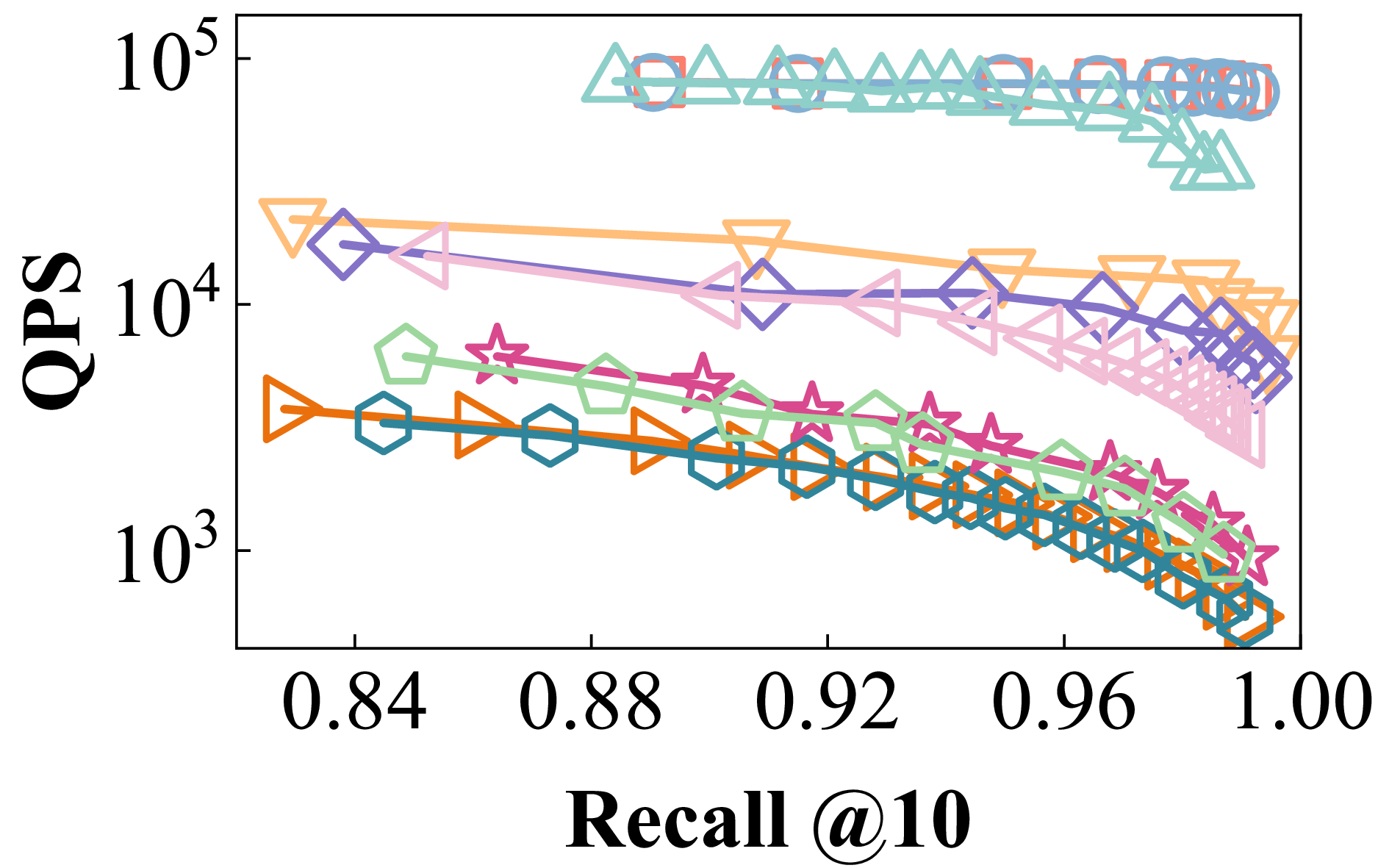}}
    \hspace{-2.2mm}
    \subfigure[Color]{
    \includegraphics[width=0.2\textwidth]{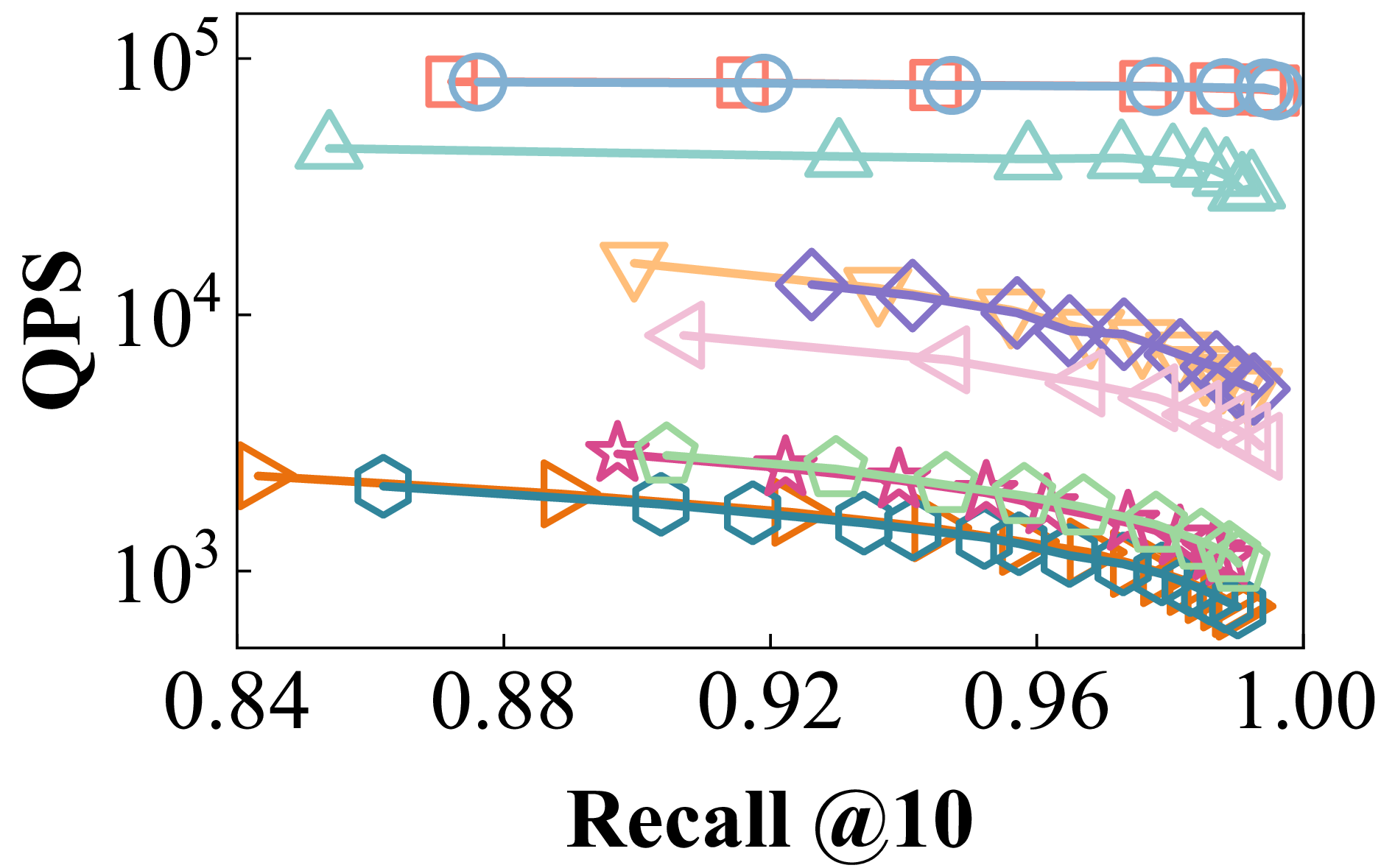}}
    \hspace{-2.2mm}
    \subfigure[Gist]{
    \includegraphics[width=0.2\textwidth]{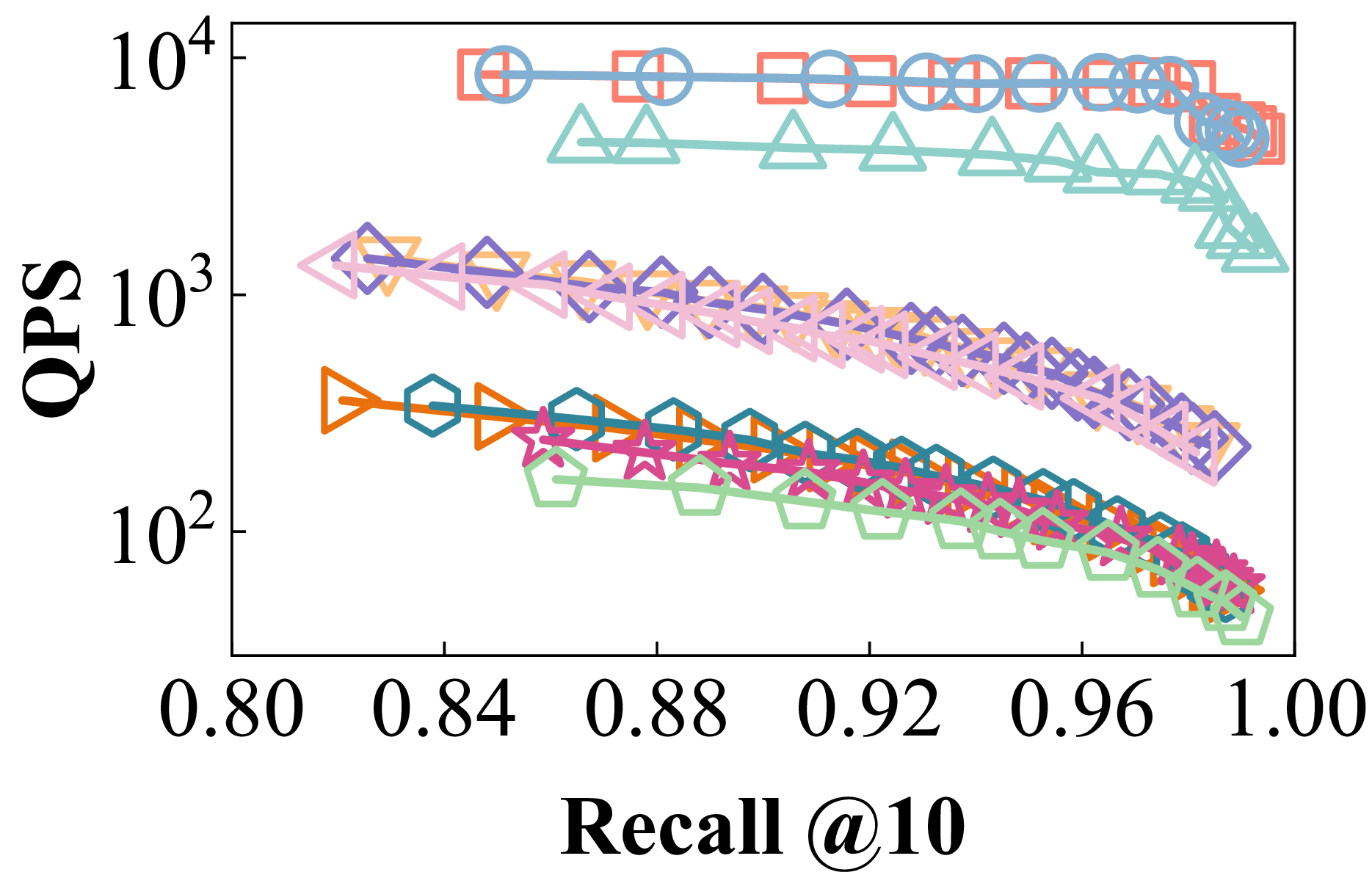}}
    \vspace{-5mm}
    \caption{Index quality on million-scale datasets.
    }
    \vspace{-3mm}
    \label{fig:quality}
\end{figure*}

\noindent\textbf{Methods and parameters.}
We evaluate our method against 7 open-sourced graph-based baselines on GPU and CPU platforms. The experimental settings, including graph degrees, thresholds (as per original papers), thread counts for construction and query, and platforms (CPU or GPU), are summarized in Table \ref{tab:exp_setting}. For fair comparison, we fix the number of CPU threads for index construction and query at 20 and 1, respectively.
The compared baselines are listed as follows.

\noindent (1) \textbf{CAGRA}~\cite{ootomo2024cagra} is the state-of-the-art graph indexing method accelerated by GPU. 
{\sf \textbf{Tagore}}\textbf{-CAGRA} optimizes CAGRA with {\sf Tagore}. 
We employ the query approach in CAGRA using the GPU to evaluate the quality of indexes constructed by both methods. 

\noindent (2) \textbf{GANNS}~\cite{yu2022gpu} is the state-of-the-art increment-based method that accelerates the construction of HNSW using the GPU. 

\noindent (3) \textbf{SymQC}~\cite{gou2025symphonyqg} is the state-of-the-art graph indexing method accelerated by quantization and SIMD instruction on CPUs. It randomly initializes a graph, iteratively collects candidates for every node, and employs the pruning strategy in NSG~\cite{fu2017nsg} to filter the candidates. 
For a fair comparison, we employ the query approach in NSG instead of that in SymQC with quantization optimization to evaluate the quality of indexes. 

\noindent (4) \textbf{NSG}~\cite{fu2017nsg} is a CPU-based method employing distance-based pruning strategy. {\sf \textbf{Tagore}}\textbf{-NSG} is its optimized version in {\sf Tagore}. 

\noindent (5) \textbf{NSSG}~\cite{fu2021high} is a refinement-based method adopting angle-based pruning strategy. {\sf \textbf{Tagore}}\textbf{-NSSG} is a variant accelerated by {\sf Tagore}. 

\noindent (6) \textbf{DPG}~\cite{li2019approximate} is another refinement-based method employing angle-based pruning strategy. {\sf \textbf{Tagore}}\textbf{-DPG} optimizes this method on the GPU. 

\noindent (7) \textbf{Vamana}~\cite{jayaram2019diskann} is designed for indexing large-scale datasets integrated in DiskANN. {\sf \textbf{Tagore}}\textbf{-Vamana} optimizes this method on the GPU-CPU-disk asynchronous framework. We compare both methods on the two billion-scale datasets in Table \ref{tab:dataset}. The pruning parameter $\alpha$ is set to 1.2, and the maximal degree is 32. For DiskANN, both datasets are partitioned into 40 clusters with an average size of 50M, while 400 clusters with an average size of 5M for {\sf Tagore} due to the limited GPU memory capacity. Moreover, we allocate 24 GB of memory to store the local indexes on the CPU for both methods. 

\begin{table}[tbp]
\renewcommand{\arraystretch}{0.95}
    \centering
    \small
    \caption{Details of Datasets. $N$ and $N_q$ denote the number of data and query vectors. $D$ denotes the vector dimension. }
    \vspace{-3mm}
    \begin{tabular}{p{2.0cm}p{1.35cm}<{\centering}p{1.0cm}<{\centering}p{1.0cm}<{\centering}p{1.4cm}<{\centering} }
        \toprule
         \textbf{{Datasets}} & $\mathbf{N}$ & $\mathbf{D}$ & \textbf{Size(GB)} & $\mathbf{N_q}$ \\
         \midrule
         Deep-1M~\cite{babenko2016efficient} & 1,000,000 & 96 & 0.36 & 10,000 \\
         SIFT~\cite{sift_and_gist} & 1,000,000 & 128 & 0.48 & 10,000 \\
         UKBench~\cite{nister2006scalable} & 1,097,907 & 128 & 0.52 & 200 \\
         Color~\cite{color_dataset} & 1,000,000 & 282 & 1.05 & 10,000 \\
         Gist~\cite{sift_and_gist} & 1,000,000 & 960 & 3.58 & 1,000 \\
         BIGANN~\cite{sift_and_gist} & 1,000,000,000 & 128 & 119.21 & 10,000 \\
         Deep-1B~\cite{babenko2016efficient} & 1,000,000,000 & 96 & 357.63 & 10,000 \\
        \bottomrule
    \end{tabular}
    \label{tab:dataset}
    \vspace{-2mm}
\end{table}

\begin{table}[tbp]
\renewcommand{\arraystretch}{0.95}
    \centering
    \small
    \caption{{Experimental settings of the compared algorithms. Parm. is short for Parameters, Constr. is short for Construction, and $t$ in CPU($t$) denotes the number of threads.} }
    \vspace{-3mm}
    \begin{tabular}{p{6mm}<{\centering}p{7mm}<{\centering}p{7mm}<{\centering}p{7mm}<{\centering}p{7mm}<{\centering}p{7.5mm}<{\centering}p{7.5mm}<{\centering}p{8.5mm}<{\centering}p{12mm}<{\centering} }
         \toprule
         Parm. & CAGRA & GANNS & SymQC & NSG & NSSG & DPG & Vamana\\
        \toprule
        Degree & $=32$ & $[32, 64]$ & $=32$ & $(0, 50]$ & $(0, 50]$ & $=32$ & $(0, 32]$ \\
        Constr. & GPU & GPU & CPU(20) & CPU(20) & CPU(20) & CPU(20) & CPU(20)\\
        Query & GPU & GPU & CPU(1) & CPU(1) & CPU(1) & CPU(1) & CPU(1)\\
        Others & - & - & - & - & $\gamma=60$ & - & $\alpha=1.2$ \\
    \bottomrule
    \end{tabular}
    \label{tab:exp_setting}
    \vspace{-2mm}
\end{table}

\noindent\textbf{Metrics.} To quantify the index construction time of the benchmarked methods, we measure their index construction time. Index quality is evaluated via two metrics: \textit{Queries Per Second (QPS)} and \textit{Recall}, which reflect the query efficiency and accuracy, respectively. \textit{QPS} is defined as the number of queries processed per second, while \textit{Recall} is calculated by $\frac{|G\cap R|}{k}$, where $G$ is the ground-truth top-$k$ nearest neighbors and $R$ is the search results. 

\noindent\textbf{Platforms.} All experiments are conducted on an Ubuntu 22.04 server, featuring an Intel Xeon Silver 4316 CPU@2.30GHz, 251GB RAM, and a Nvidia GeForce RTX 4090 GPU (24G). We implement {\sf Tagore} in C++/CUDA under CUDA 12.2 and integrate it into Python. 

\subsection{Overall Indexing Performance}
\label{subsec:overall_performance}

\subsubsection{\textbf{Indexing performance on million-scale datasets.}} 
Figure \ref{fig:over_performance} summarizes the index construction time across the evaluated methods. All refinement-based indexes accelerated by {\sf Tagore} demonstrate superior performance, outperforming the baselines by significant margins. Compared to GPU-based methods, {\sf Tagore}-CAGRA achieves speedups of 1.32$\times$ to 6.39$\times$ over CAGRA and 1.99$\times$ to 10.14$\times$ over GANNS, which highlights the superiority of our designed GPU-specific algorithms (e.g., GNN-descent and parallel filtering kernels). 
Against CPU-based SymQC, which leverages SIMD optimizations, {\sf Tagore}-NSG attains 11.74$\times$ to 47.73$\times$ faster construction. Furthermore, {\sf Tagore} accelerates NSG, NSSG, and DPG by 36.23$\times$ to 64.42$\times$, 43.51$\times$ to 112.79$\times$, and 21.10$\times$ to 51.31$\times$, respectively.  
While all four {\sf Tagore}-enhanced methods exhibit comparable indexing times, {\sf Tagore}-CAGRA shows marginal efficiency gains due to its GPU-friendly and distance-free pruning strategy. 
On average, {\sf Tagore}-CAGRA achieves a 4.65$\times$ speedup over the state-of-the-art GPU-based method CAGRA, which can be attributed to the rapid convergence of GNN-Descent and the balanced workload achieved by the balanced-parallel model. 
Notably, in Figure \ref{fig:over_performance}, the index construction time of {\sf Tagore} increases on dataset Gist, which has an extremely high dimension of 960, as our optimizations focus on algorithmic improvements rather than low-level distance computation enhancements. The distance calculation optimizations (e.g., using PTX assembly instructions) remain as future work. 

To evaluate the quality of indexes constructed using {\sf Tagore} versus baseline methods, we measure query performance (accuracy and efficiency) using each method's native query implementation. The query performance is exhibited in Figure \ref{fig:quality}. 
Please note that the query methods are consistent within the baseline indexing method and the corresponding methods in {\sf Tagore} (e.g., NSG vs. {\sf Tagore}-NSG) but differ across different types of indexes (see Section \ref{subsec:setting}). Thus, results can only be used to compare the quality between indexes constructed by the baseline method and the corresponding method in {\sf Tagore} (e.g., NSG and {\sf Tagore}-NSG) but not across distinct index types (e.g., NSG vs. NSSG). As shown in Figure \ref{fig:quality}, the indexes constructed by {\sf Tagore} achieve comparable recall and QPS to their baseline counterparts, which demonstrates that {\sf Tagore} accelerates construction without degrading index quality. 

\subsubsection{\textbf{Indexing performance on billion-scale datasets.}} 
\begin{figure}
    \centering
    \includegraphics[width=0.48\textwidth]{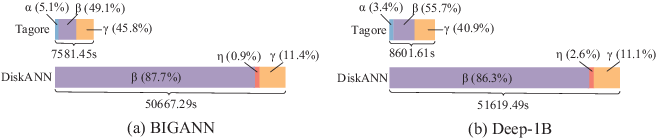}
    \vspace{-6mm}
    \caption{Index construction time on billion-scale datasets. $\alpha$ denotes the data transfer overhead from CPU to GPU, $\beta$ denotes the local index construction overhead, $\eta$ denotes the writing overhead of local index, and $\gamma$ denotes the local index merging overhead to the disk. }
    \label{fig:billion_performance}
    \vspace{-2mm}
\end{figure}

Figure \ref{fig:billion_performance} compares the indexing performance of {\sf Tagore} and DiskANN~\cite{jayaram2019diskann} on billion-scale datasets. Overall, the indexing time of {\sf Tagore} is 6.68$\times$ and 6.00$\times$ less than that of DiskANN on datasets BIGANN and Deep-1B, respectively. {\sf Tagore} reduces the index construction time of billion-scale datasets from 14 hours to under 3 hours using a single GPU, meeting nightly index rebuild requirements. 
As shown in the time breakdown (Figure \ref{fig:billion_performance}), the data transfer overhead from CPU to GPU is insignificant, accounting for only 3\%-5\% of the whole index construction time. Different from the query process, indexing requires infrequent data exchange between the CPU and GPU. Consequently, the GPU is a viable accelerator for index construction, featuring low data movement overhead and high index construction efficiency. 
In DiskANN, local index construction is the most time-consuming procedure during the index construction of billion-scale datasets, which demands substantial calculations. In contrast, this bottleneck is well overcome by the GPU in {\sf Tagore}. {\sf Tagore} improves the local index construction by 11.9$\times$ and 9.3$\times$ over DiskANN on BIGANN and Deep-1B, respectively. 
Moreover, while merging local indexes into the global index is not the bottleneck in DiskANN, it becomes the bottleneck in {\sf Tagore} due to the efficient construction of local indexes. The cluster-aware caching mechanism proposed in Section \ref{subsec:caching} plays a critical role in reducing the merging overhead, which will be further analyzed in Section \ref{subsec:eva_caching}.

To further improve computational efficiency and scalability, {\sf Tagore} supports index construction with multiple GPUs. Since multi-GPU index construction doesn't impact the merging phase, we only report the overhead of local index construction in Table \ref{tab:multiGPU}. Results demonstrate {\sf Tagore}'s ability of near-linear scaling: the construction time decreases linearly with the increasing number of GPUs. With 4 GPUs, {\sf Tagore} reduces local index construction overhead by 47.7$\times$ and 36.17$\times$ on BIGANN and Deep-1B compared to DiskANN, while the speedups are 24.0$\times$ and 18.1$\times$ with 2 GPUs. 

\begin{figure}
    \centering    
    \includegraphics[width=0.28\textwidth]{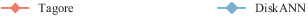}\\
     \vspace{-2mm}
    \subfigure[BIGANN]{
    \includegraphics[width=0.23\textwidth]{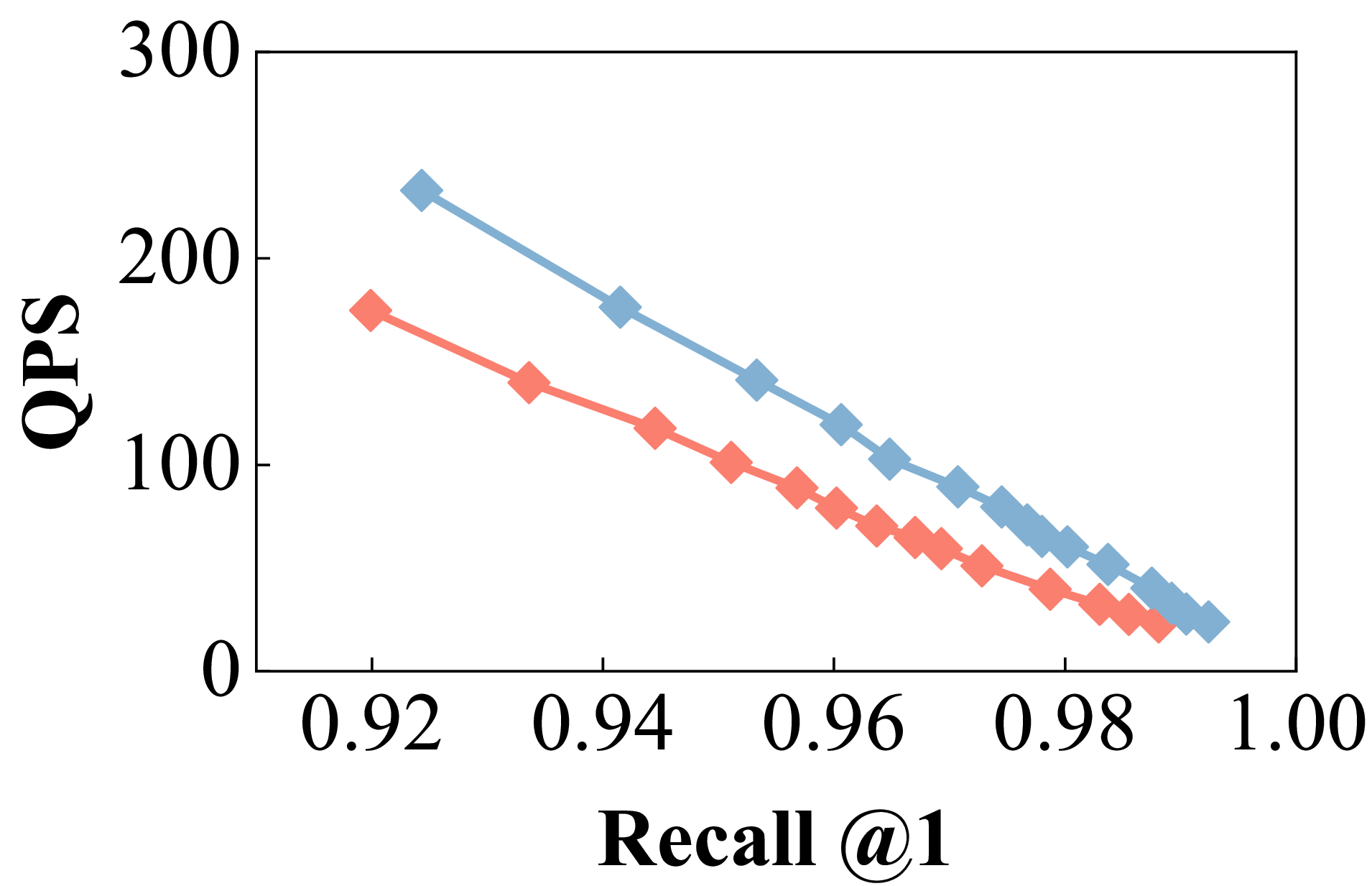}}
    \subfigure[Deep-1B]{
    \includegraphics[width=0.22\textwidth]{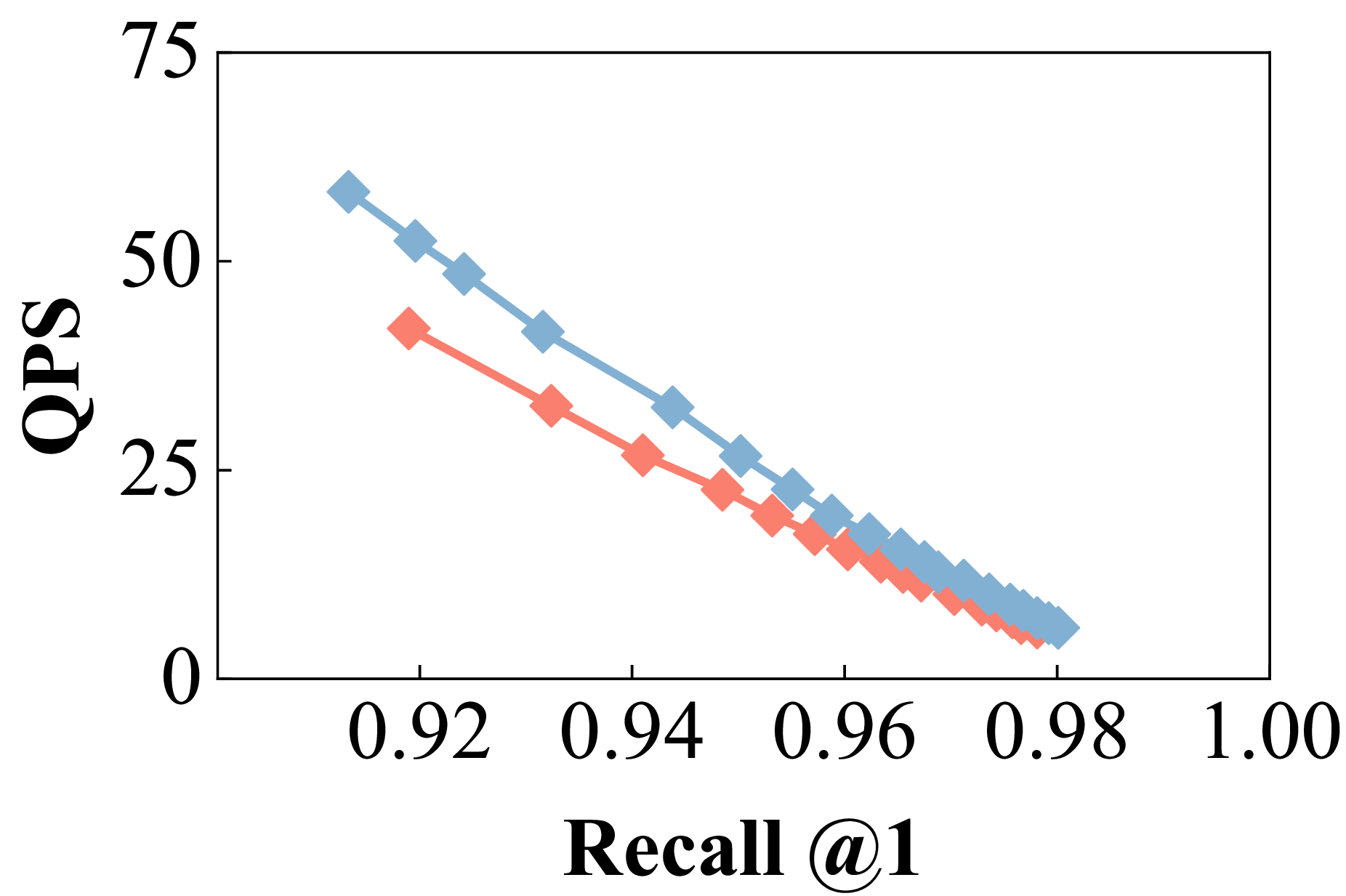}}
    \vspace{-3mm}
    \caption{Index quality on billion-scale datasets.}
    \label{fig:big_quality}
    \vspace{-3mm}
\end{figure}

We evaluate the global index quality using the query method in DiskANN on a Samsung 980 SSD (1 TB). The query performance on both datasets is depicted in Figure \ref{fig:big_quality}. 
When the recall is greater than 95\%, the difference in recall between {\sf Tagore} and DiskANN under the same QPS is only within 1.4\%. When the recall is greater than 98\%, this gap narrows down to 0.5\%, which is tolerable. The index quality constructed by {\sf Tagore} is slightly lower than that of DiskANN because {\sf Tagore} partitions the datasets into 400 clusters due to the limitation of GPU memory, while DiskANN partitions them into 40 clusters. More clusters are harmful to the global neighbor relationships of the index. 

\begin{table}[tbp]
\renewcommand{\arraystretch}{0.95}
    \centering
    \small
    \caption{Local index construction overhead with multiple GPUs. $a$ in {\sf Tagore}-$a$ denotes the number of GPUs. }
    \vspace{-3mm}
    \begin{tabular}{p{1.2cm}<{\centering}p{1.4cm}<{\centering}p{1.4cm}<{\centering}p{1.4cm}<{\centering}p{1.4cm}<{\centering} }
         \toprule
         Datasets & DiskANN & {\sf Tagore}-1 & {\sf Tagore}-2 & {\sf Tagore}-4\\
        \toprule
        BIGANN & 44456.13s & 3721.08s & 1850.90s & \textbf{933.65s} \\
        Deep-1B & 44538.57s & 4791.77s & 2454.39s & \textbf{1231.46s} \\
    \bottomrule
    \end{tabular}
    \label{tab:multiGPU}
    \vspace{-4mm}
\end{table}

\subsection{Evaluation of GNN-Descent}
\label{subsec:eva_nndescent}

\begin{figure}
    \centering    
    \hspace{3mm}
    \includegraphics[width=0.48\textwidth]{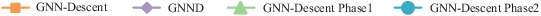}\\
    \vspace{-2mm}
    \subfigure[SIFT]{
    \includegraphics[width=0.22\textwidth]{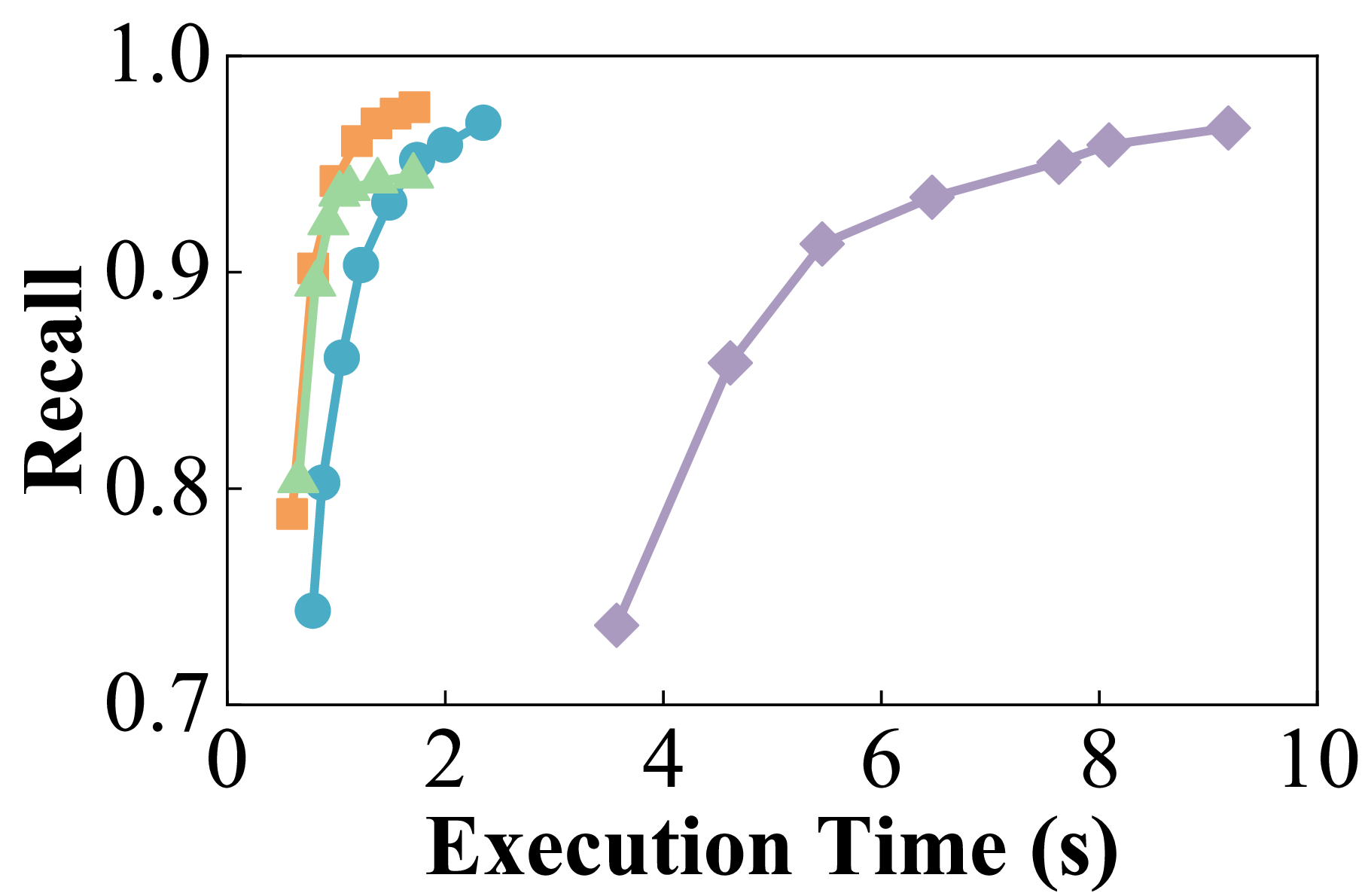}}
    \subfigure[Color]{
    \includegraphics[width=0.22\textwidth]{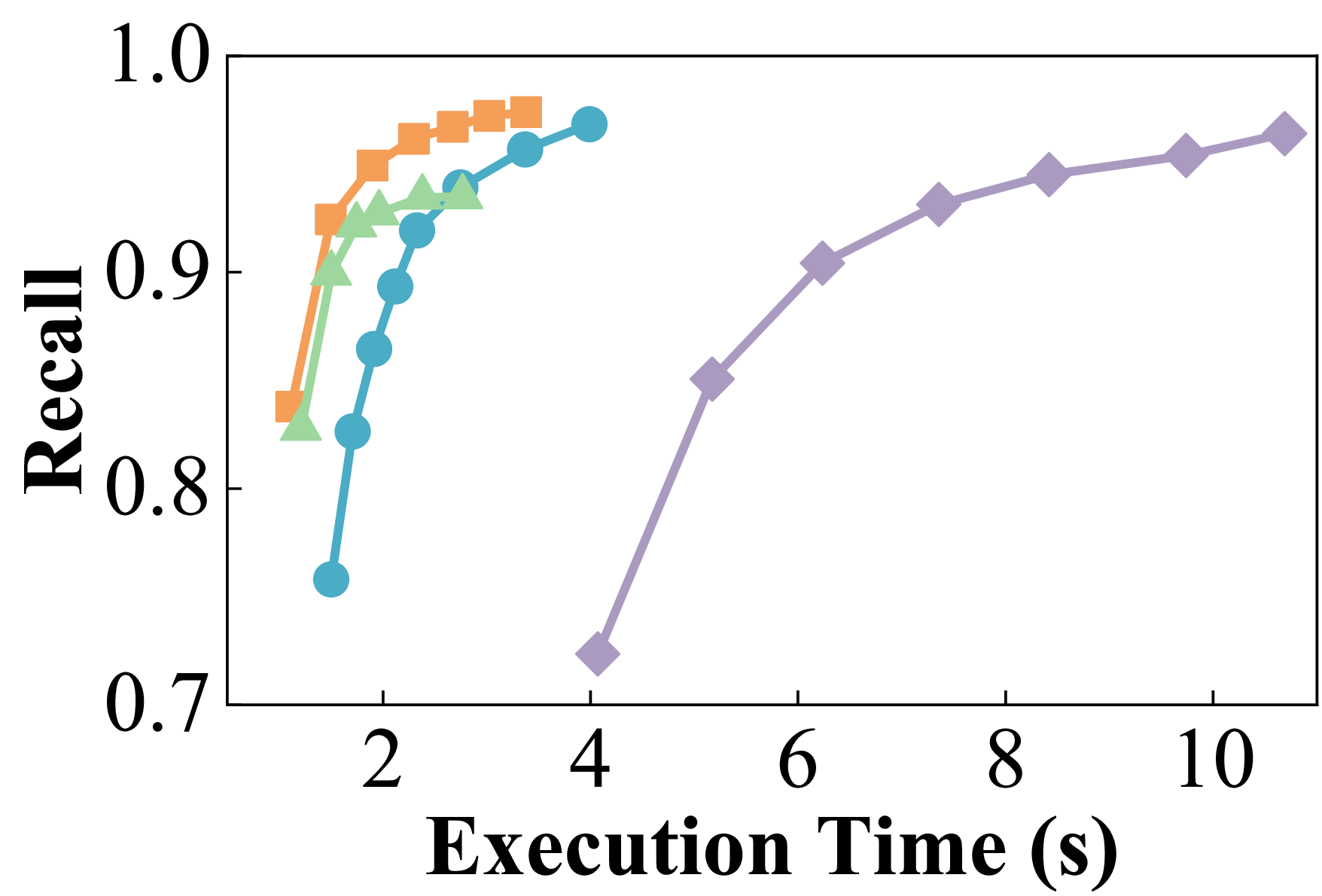}}
    \vspace{-3mm}
    \caption{Performance comparison of NN-Descent on GPU. 
    }
    \label{fig:nn_descent_compare}
    \vspace{-3mm}
\end{figure}

To evaluate the GNN-Descent algorithm proposed in Section \ref{sec:gnn_descent}, we benchmark GNN-Descent against the state-of-the-art GPU-based NN-Descent algorithm, GNND~\cite{wang2021fast}, implemented in cuVS~\cite{cuvs2024}. Figure \ref{fig:nn_descent_compare} compares their performance on two representative datasets, including two single-phase variants of GNN-Descent. 
GNN-Descent achieves speedups of 6.7$\times$ and 4.0$\times$ over GNND on the respective datasets. 
As observed in Figure \ref{fig:nn_descent_compare}, the first phase of GNN-Descent converges rapidly compared to the second phase in the initial iterations. Its convergence speed slows near 93\% recall and follows a similar trend to GNND. In contrast, the second phase sustains high convergence rates, demonstrating the effectiveness of our theory proposed in Section~\ref{sec:gnn_descent}. 
With the fine-grained sampling in the second phase, GNN-Descent can maintain a high convergence rate even at high recall (>95\%), significantly enhancing the efficiency of NN-Descent on GPUs. 
Furthermore, although neither single-phase variant outperforms GNN-Descent, both are superior to GNND. This is because we offload all the operations of NN-Descent onto the GPU, eliminating the overhead of CPU-GPU synchronization and data transfer. Moreover, we optimize the neighbor updates by proposing a lock-free merging strategy, effectively improving the efficiency of neighbor updates in GNN-Descent. 

\begin{figure}
    \centering    
    \includegraphics[width=0.38\textwidth]{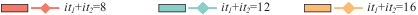}\\
    \vspace{-2mm}
    \subfigure[SIFT]{
    \includegraphics[width=0.23\textwidth]{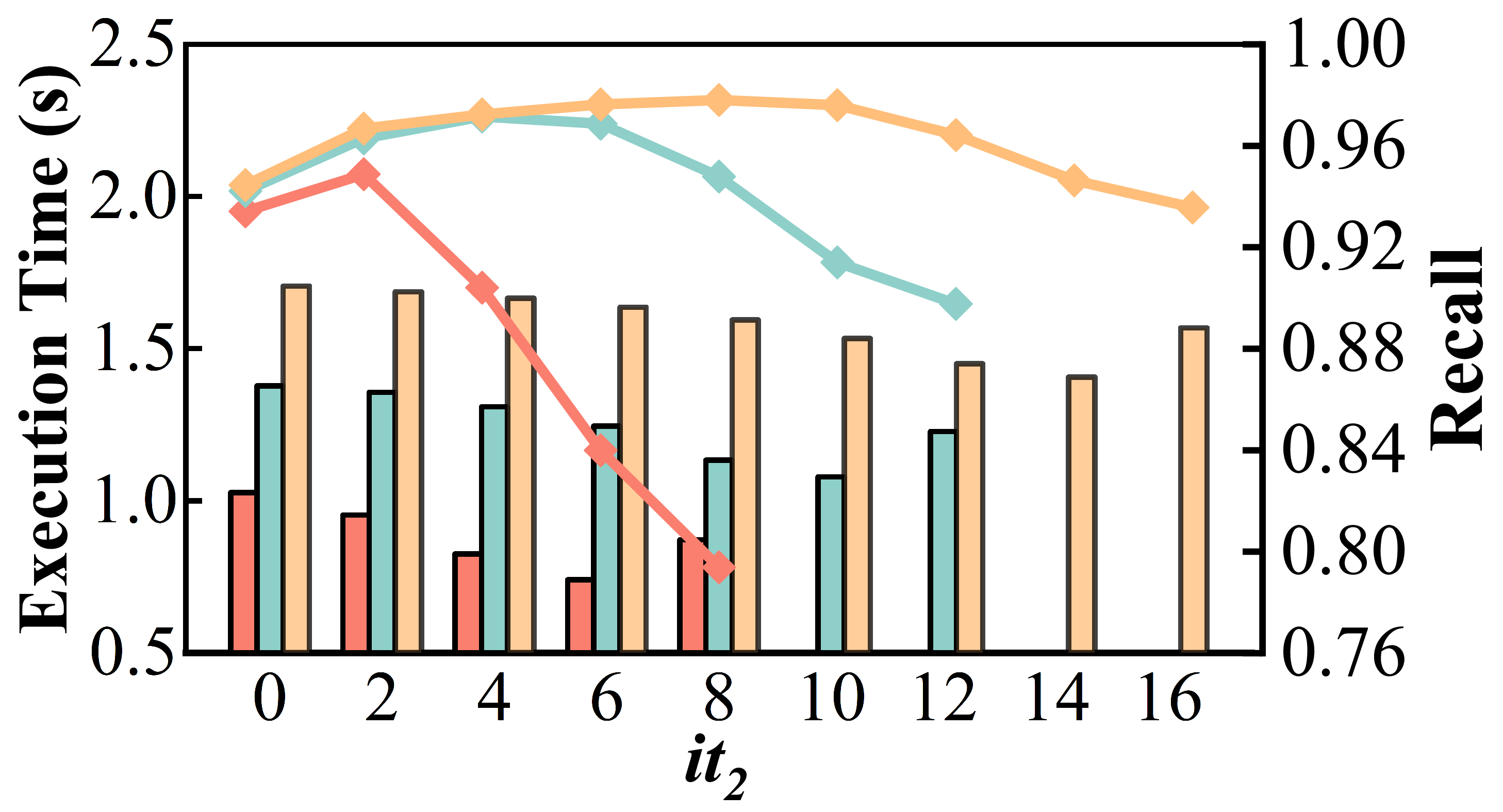}}
    \subfigure[Color]{
    \includegraphics[width=0.23\textwidth]{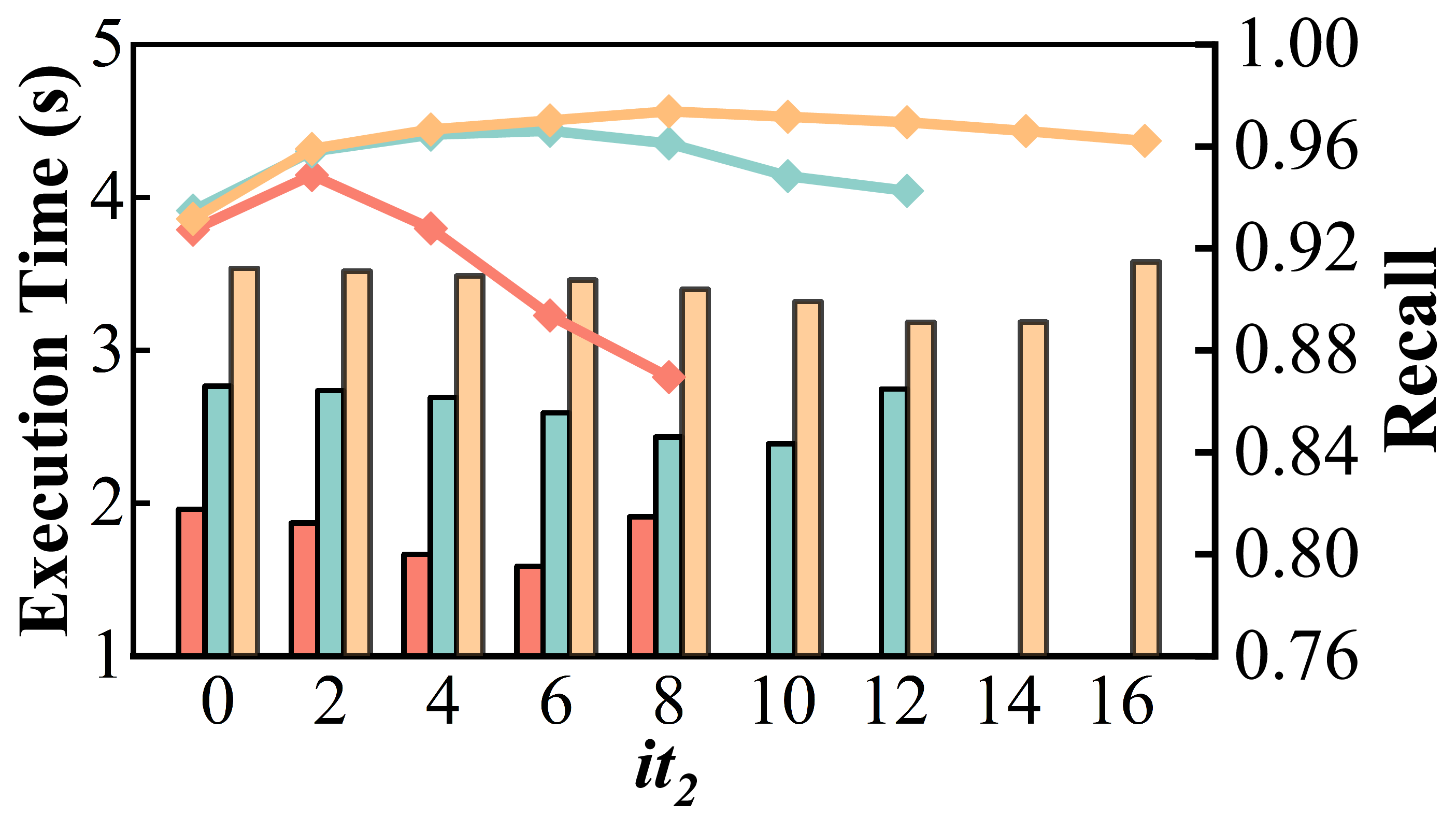}}
    \vspace{-5mm}
    \caption{Impacts of iteration counts $it_1$ and $it_2$ on the effectiveness of on GNN-Descent. Bars and lines represent execution time and recall, respectively. }
    \label{fig:nn_descent_parameter}
    \vspace{-3mm}
\end{figure}

In GNN-Descent (Algorithm \ref{alg:two_phase_descent}), the parameters $it_1$ and $it_2$ define the number of iterations required in two distinct phases.  While selecting an appropriate total number of iterations  ($it_1+it_2$) is relatively straightforward -- users can adjust it according to their desired recall -- the difficulty lies in determining how to allocate iterations between the two phases. To explore the impact of different allocation strategies, we evaluate the performance of GNN-Descent by keeping the total iteration count $it_1+it_2$ fixed and varying the distribution between $it_1$ and $it_2$ on two representative datasets.
As shown in Figure~\ref{fig:nn_descent_parameter}, when the total iteration count is fixed at 8, 12, or 16, the recall on both datasets increases initially and then decreases, while the execution time exhibits an opposite trend, as we increase $it_2$ (and decrease $it_1$). 
This aligns with the observation in Figure~\ref{fig:nn_descent_compare}, where the recall achieved during Phase 1 plateaus after reaching a high value, even if the execution time is further extended. 
This occurs because, when $it_1$ is large, the later iterations in the first phase involve many distance calculations but contribute only minimal neighbor updates.  
Introducing more iterations in the second phase enables GNN-Descent to continue refining neighbors in a more fine-grained manner, thereby improving the recall value.
However, if $it_1$ becomes significantly smaller than $it_2$, the algorithm transitions to the second phase before sufficient recall has been achieved in the first phase.
{The execution time of each Phase 2 iteration varies, as more neighbor updates lead to longer execution times. As shown in Figure \ref{fig:nn_descent_compare}, Phase 2 is less effective at improving index quality when recall is low. As $it_2$ increases, the earlier iterations yield poor index quality, causing more neighbors to be updated in later iterations. This results in a higher execution time for the final iterations, which consequently increases the overall execution time. And further increasing $it_2$ beyond a certain point leads to reduced recall. }
To balance efficiency and accuracy, we set $it_1=it_2$ in all experiments. This configuration provides a favorable trade-off, achieving strong performance with low computational overhead. 

\subsection{Evaluation of Pruning Optimizations}
\label{subsec:eva_pruning}

To assess the efficacy of our proposed pruning optimizations, we benchmark {\sf Tagore} against the original pruning strategies listed in Table \ref{tab:prun_summary}. The results are illustrated in Table~\ref{tab:pruning}. {\sf Tagore} achieves speedups of 16.57$\times$ to 19.43$\times$, 74.36$\times$ to 127.95$\times$, 179.04$\times$ to 342.48$\times$, 22.15$\times$ to 27.94$\times$ over the pruning phase of CAGRA, NSG, NSSG, and DPG, respectively. As mentioned in Section \ref{sec:pruning}, although CAGRA performs the pruning procedure on the GPU, it has not fully utilized the parallelism of the GPU and hence suffers a severely unbalanced workload. Our proposed balanced-parallel computing model mitigates these issues, delivering acceleration of over an order of magnitude. For NSG, NSSG, and DPG, which rely heavily on incremental and compute-intensive distance calculations, our proposed wavefront parallelism strategy transforms the incremental distance calculations into batched GPU-friendly tasks, enabling up to two orders of magnitude speedups. Moreover, we also implement a GPU-based serial computing paradigm in Figure \ref{fig:incrental}(a), which exhibits a 1.9$\times$ performance gap on average compared with the wavefront parallelism strategy in Figure~\ref{fig:incrental}(b). 
Notably, in Table~\ref{tab:pruning}, the speedup ratio achieved by {\sf Tagore} is more significant on datasets with higher dimensions. This is attributed to the powerful computing capabilities of GPUs that significantly accelerate distance calculations, making them more suitable for constructing indexes for high-dimensional datasets compared to CPUs. 

\begin{table}[tbp]
\renewcommand{\arraystretch}{0.95}
\belowrulesep=0pt
\aboverulesep=0pt
    \centering
    \small
    \caption{Pruning overhead of the refinement-based methods (unit: second). O denotes that of the original methods and T denotes the execution time of {\sf Tagore}. }
    \vspace{-3mm}
    \begin{tabular}{p{1.3cm}<{\centering}|p{0.5cm}<{\centering}p{0.5cm}<{\centering}|p{0.4cm}<{\centering}p{0.6cm}<{\centering}|p{0.4cm}<{\centering}p{0.6cm}<{\centering}|p{0.5cm}<{\centering}p{0.5cm}<{\centering} }
         \toprule
    \multirow{2}{*}{Dataset} & \multicolumn{2}{c|}{CAGRA} & \multicolumn{2}{c|}{NSG} & \multicolumn{2}{c|}{NSSG} & \multicolumn{2}{c}{DPG} \\
    \cline{2-9}
      & O & T & O & T & O & T & O & T \\
      \midrule
      Deep-1M & 1.33 & \textbf{0.07} & 39.55 & \textbf{0.51} & 80.57 & \textbf{0.45} & 5.98 & \textbf{0.27} \\
      SIFT & 1.16 & \textbf{0.07} & 37.18 & \textbf{0.50} & 102.30 & \textbf{0.48} & 6.87 & \textbf{0.27} \\
      UKBench & 1.43 & \textbf{0.08} & 37.55 & \textbf{0.50} & 103.37 & \textbf{0.47} & 7.39 & \textbf{0.30} \\
      Color & 1.20 & \textbf{0.07} & 111.61 & \textbf{0.98} & 184.94 & \textbf{0.54} & 13.41 & \textbf{0.48} \\
      Gist & 1.36 & \textbf{0.07} & 239.26 & \textbf{1.87} & 323.22 & \textbf{1.66} & 37.32 & \textbf{1.45} \\
    \bottomrule
    \end{tabular}
    \label{tab:pruning}
    \vspace{-2mm}
\end{table}

\subsection{Evaluation of Caching Mechanism}
\label{subsec:eva_caching}

In DiskANN, the local indexes are first written to the disk and then merged,  requiring three CPU-disk I/O operations per node. 
As depicted in Figure~\ref{fig:billion_performance}, the writing and merging overhead of DiskANN is 6211.16s and 7080.92s on the BIGANN and Deep-1B datasets, respectively. 
While this overhead is secondary to DiskANN's dominant local indexing costs, it becomes the main bottleneck in Tagore, where I/O overhead surpasses local indexing costs. 
To address this issue, {\sf Tagore} merges the neighbor lists of nodes in the cache directly, eliminating the need for writing to disk first and
reducing the disk I/O overhead. Moreover, {\sf Tagore} pipelines the merging procedure on the CPU with local indexing on the GPU. 
Consequently, the merging overhead is reduced to 3475.95s and 3521.04s for the two billion-scale datasets, achieving speedup ratios of 1.79$\times$ and 2.01$\times$, respectively, demonstrating the effectiveness of our proposed cluster-based caching mechanism. 

\begin{table}[tbp]
\renewcommand{\arraystretch}{0.95}
    \centering
    \small
    \caption{Comparison of cache hit ratio between various cluster dispatching orders. }
    \vspace{-3mm}
    \begin{tabular}{p{1.2cm}<{\centering}p{1.4cm}<{\centering}p{1.4cm}<{\centering}p{1.4cm}<{\centering}p{1.4cm}<{\centering} }
         \toprule
         Datasets & Sequential & Random & Gorder & {\sf Tagore}\\
        \toprule
        BIGANN & 13.1\% & 13.6\% & 24.9\% & \textbf{61.6\%} \\
        Deep-1B & 12.1\% & 12.5\% & 19.1\% & \textbf{67.2\%} \\
    \bottomrule
    \end{tabular}
    \label{tab:cache}
    \vspace{-2mm}
\end{table}

The key innovation of our cluster-aware caching mechanism is the cluster dispatching order in Algorithm \ref{alg:cluster_order}. To demonstrate its superior performance, we compare it against three baselines: sequential order, random order, and a state-of-the-art graph reorder algorithm Gorder~\cite{wei2016speedup}. Gorder generates a permutation of all nodes in a graph based on node locality to minimize CPU cache misses. We apply Gorder to the cluster graph depicted in Figure \ref{fig:cache_mechanism}(b). 
Table \ref{tab:cache} presents the cache hit ratio for various cluster dispatching orders. 
The sequential order, which dispatches the clusters in their original sequence, yields the lowest cache hit ratios of 13.1\% and 12.1\%. 
Random cluster dispatching shows a marginal improvement over sequential order. 
Gorder enhances the locality of clusters in the cache, leading to a significant improvement in the hit rate. However, it still has limitations. 
The cluster dispatching order in {\sf Tagore} achieves hit rates of 61.6\% (BIGANN) and 67.2\% (Deep-1B), which are 2.47$\times$ and 3.52$\times$ higher than those of Gorder. 
Furthermore, {\sf Tagore} generates the dispatching order before index construction, with overheads of just 1.27s and 1.25s on the two datasets, which is negligible compared to the total index construction time. 
\section{Conclusions}
\label{sec:conclusion}
In this paper, we introduce {\sf Tagore}, a scalable GPU-accelerated library for efficient graph indexing, specializing in refinement-based methods. 
By harnessing GPU parallelism, {\sf Tagore} optimizes the indexing pipeline with three innovations.
Firstly, we propose a two-phase algorithm, GNN-Descent, to achieve efficient \textit{k}-NN graph initialization, accelerating convergence through adaptive sampling criteria. A lock-free neighbor list update algorithm is designed for GNN-Descent to unleash the parallel potential of GPUs. 
Secondly, we present a unified pipeline CFS to accelerate the distinct refinement-based strategies and integrate two GPU-tailored kernels for diverse pruning strategies. 
Thirdly, we propose a hybrid GPU-CPU-disk asynchronous framework augmented by a cluster-aware caching mechanism to efficiently construct billion-scale indexes, minimizing the I/O bottlenecks. 
Experimental results consistently demonstrate the superiority of {\sf Tagore}. 

For future work, there remain two critical challenges. 
First, during \textit{k}-NN graph initialization, late-stage iterations incur significant overhead from redundant distance calculations. Most sampled nodes in later phases lie far outside the current node’s neighborhood, rendering brute-force distance comparisons wasteful. A probabilistic filtering mechanism that skips low-likelihood candidates could replace brute-force distance computations, reducing GPU workload.
Second, as index construction has been accelerated by {\sf Tagore}, disk I/O (data loading, merging) becomes the dominant bottleneck. Integrating NVMe SSDs and CXL memory pools could mitigate this by enabling direct data streaming between GPUs and high-bandwidth storage/memory tiers. 

\begin{acks}
This work was supported in part by the NSFC under Grants No. (62025206, U23A20296), Zhejiang Province’s “Lingyan” R\&D Project under Grant No. 2024C01259, CCF-Aliyun2024004, and Yongjiang Talent Introduction Programme (2022A-237-G). Yunjun Gao is the corresponding author of the work.
\end{acks}

\bibliographystyle{ACM-Reference-Format}
\bibliography{sample-base}


\end{document}